\documentclass[letterpaper,journal]{IEEEtran}
\usepackage{amsmath,amsfonts, amssymb}
\usepackage{array}
\usepackage{booktabs}
\usepackage{textcomp}
\usepackage{stfloats}
\usepackage{url}
\usepackage{tabularx}
\usepackage{verbatim}
\usepackage{graphicx}
\usepackage{svg}
\usepackage[hidelinks,unicode]{hyperref} 
\hypersetup{
    pdftitle={
        Talking Ink: A Flow-Based Multi-Molecule Molecular Communication
        Testbed for Effective Channel Modeling and Detector Benchmarking
    },
    pdfauthor={
        Alexander Wietfeld, Krupakar Duvva, Jiachen Xiao, Wolfgang Kellerer
    },
    pdfsubject={
        Experimental molecular communication testbed, effective channel
        modeling, and detector benchmarking
    },
    pdfkeywords={
        channel modeling, detection schemes, molecular communication,
        multi-molecule transmission, spectral sensing, testbeds
    },
    pdflang={en-US},
    pdfdisplaydoctitle=true,
    bookmarksnumbered=true,
    bookmarksopen=false
}

\usepackage[acronym,shortcuts,nonumberlist]{glossaries} 
\newacronym{DBMC}{DBMC}{diffusion-based molecular communication}
\newacronym{TDMA}{TDMA}{time-division multiple access}
\newacronym{MDMA}{MDMA}{molecule-division multiple access}
\newacronym{NOMA}{NOMA}{non-orthogonal multiple access}
\newacronym{ADMA}{ADMA}{amplitude-division multiple access}
\newacronym{MA}{MA}{multiple access}
\newacronym{BEP}{BEP}{bit error probability}
\newacronym{TX}{TX}{transmitter}
\newacronym{RX}{RX}{receiver}
\newacronym{PCB}{PCB}{printed circuit board}
\newacronym{ISI}{ISI}{inter-symbol interference}
\newacronym{MAI}{MAI}{multiple-access interference}
\newacronym{SIC}{SIC}{successive interference cancellation}
\newacronym{OOK}{OOK}{on--off keying}
\newacronym{MCS}{MCS}{Monte Carlo simulation}
\newacronym{SNR}{SNR}{signaling-molecule-to-noise ratio}
\newacronym{MC}{MC}{molecular communication}
\newacronym{IoBNT}{IoBNT}{Internet of Bio-Nano-Things}
\newacronym{BNM}{BNM}{Bio-Nano-Machine}
\newacronym{DBMC-NOMA}{DBMC-NOMA}{NOMA for DBMC networks}
\newacronym{UCA}{UCA}{uniform concentration assumption}
\newacronym{CRN}{CRN}{chemical reaction network}
\newacronym{ODE}{ODE}{ordinary differential equation}
\newacronym{SSA}{SSA}{stochastic simulation algorithm}
\newacronym{ChemSICal}{ChemSICal}{chemical reaction network for successive interference cancellation}
\newacronym{PDF}{PDF}{probability density function}
\newacronym{PMF}{PMF}{probability mass function}
\newacronym{RRC}{RRC}{reaction rate constant}
\newacronym{SCRN}{SCRN}{stochastic chemical reaction network}
\newacronym{DBMC-aNOMAly}{DBMC-aNOMAly}{asynchronous NOMA with pilot-symbol optimization protocol for DBMC}
\newacronym{EM}{EM}{electromagnetic}
\newacronym{IoT}{IoT}{Internet of Things}
\newacronym{MI}{MI}{mutual information}
\newacronym{WCAM}{WCAM}{worst-case-offset avoidance mechanism}
\newacronym{CIR}{CIR}{channel impulse response}
\newacronym{AD}{AD}{advection--diffusion}
\newacronym{BER}{BER}{bit error rate}
\newacronym{BMFTR}{BMFTR}{German Federal Ministry of Research, Technology and Space}
\newacronym{CDFRMSE}{CDFRMSE}{cumulative-distribution-function root-mean-square error}
\newacronym{DD}{DD}{difference detection}
\newacronym{DFDD}{DFDD}{decision-feedback difference detection}
\newacronym{EDD}{EDD}{energy-difference detection}
\newacronym{MF}{MF}{matched filter}
\newacronym{MLSD}{MLSD}{maximum-likelihood sequence detection}
\newacronym{MEDD}{MEDD}{modified energy-difference detection}
\newacronym{MMSE}{MMSE}{minimum mean-square error}
\newacronym{MUMO}{MUMO}{multi-molecule}
\newacronym{NRMSE}{NRMSE}{normalized root-mean-square error}
\glsdisablehyper
\usepackage{cite}
\usepackage{siunitx}
\usepackage[table]{xcolor}

\usepackage{supertabular}
\newglossarystyle{acr-2col-narrow}{%
  \renewcommand*{\glsgroupheading}[1]{}
}

\begin{document}

\title{Talking Ink: A Flow-Based Multi-Molecule Molecular Communication Testbed for Effective Channel Modeling and Detector Benchmarking

\thanks{The authors are with the Chair of Communication Networks,
Technical University of Munich, Munich, Germany. Corresponding author: Alexander Wietfeld
(e-mail: alexander.wietfeld@tum.de).}
\thanks{The authors acknowledge the financial support by the German Federal Ministry of Research, Technology and Space (BMFTR) in the program of ``Souver\"an. Digital. Vernetzt.'' Joint project 6G-life, project identification number: 16KIS2414.}}

\author{Alexander Wietfeld,~\IEEEmembership{Graduate Student Member,~IEEE}, Krupakar Duvva, Jiachen Xiao, Wolfgang Kellerer,~\IEEEmembership{Fellow,~IEEE}
}



\maketitle

\begin{abstract}
Practical \ac{MC} testbeds are essential for translating theoretical concepts toward future applications while retaining physical realism, accessibility, and experimental repeatability. In this paper, we present a flow-based \acs{MC} experimental system for real-time \ac{MUMO} transmission using cyan, magenta, and yellow inks as distinguishable molecule-like signaling carriers. Three micropumps and a multi-needle injection system release the inks into a background-flow channel, and a non-invasive spectral sensor estimates the individual received color traces at variable transmitter--receiver distances. The platform supports systematic measurements of isolated-pulse responses across different distances. We interpret these measurements as effective end-to-end \acp{CIR} shaped by finite release, flow propagation, and receiver readout, and compare two compact model families. Both reproduce much of the dominant arrival timing and pulse shape, while late-tail mismatch remains the main limitation. We connect this channel characterization to communication performance through continuous \acs{MUMO} \ac{OOK} payload measurements. The resulting retrospective benchmark with full-data parameter optimization includes trace-only and \acs{CIR}-assisted methods. We introduce \ac{MEDD}, a simple adaptive extension of standard \ac{EDD}. \acs{MEDD} outperforms all other trace-only detectors with \(\boldsymbol{34/4500=0.76\%}\) \ac{BER} and closely approaches the best overall \ac{MMSE} result of \(\boldsymbol{32/4500=0.71\%}\). After targeted parameter tuning, the detector benchmark produces zero errors over the evaluated payloads at \(\boldsymbol{6}\,\mathbf{bit}/\mathbf{s}\) over \(\boldsymbol{8}\,\mathbf{cm}\) and at \(\boldsymbol{3}\,\mathbf{bit}/\mathbf{s}\) over \(\boldsymbol{24}\,\mathbf{cm}\). Overall, the results establish the platform as a reusable experimental setup for effective channel modeling, practical detector benchmarking, and future multi-molecule networking experiments.
\end{abstract}

\glsresetall


\section{Introduction}\label{sec:introduction}

\IEEEPARstart{F}{or} environments where electromagnetic signaling is impractical or undesirable, \ac{MC} offers a promising communication paradigm by using molecules as information carriers~\cite{hiyamaMolecularCommunication2008,farsad_comprehensive_2016}. This makes \ac{MC} relevant for biomedical systems, lab-on-chip communication, and chemically interfaced sensing and actuation~\cite{akyildizInternetBioNanoThings2015,chude-okonkwoMolecularCommunicationNanonetwork2017}. A central challenge is that many \ac{MC} concepts are still evaluated primarily through analytical models or simulations, while practical links involve nonideal \ac{TX} release dynamics, fluidic transport, and imperfect calibration procedures~\cite{jamaliChannelModelingDiffusive2019}.

Experimental testbeds are therefore essential for understanding which theoretical concepts remain useful in physical systems~\cite{lotterExperimentalResearchSynthetic2023e,lotterExperimentalResearchSynthetic2023f}. Flow-based testbeds are particularly attractive because they provide repeatable operation, controllable transport, and direct access to \ac{RX} traces~\cite{farsadNovelExperimentalPlatform2017,wickeExperimentalSystemMolecular2022}. However, many existing systems focus on a single molecule type, a single \ac{RX} principle, or a specific proof-of-concept communication task. This leaves a gap for reusable platforms that combine real-time \ac{MUMO} operation, distance-dependent channel characterization, and \ac{RX} benchmarking using complementary measurements from the same experimental platform.

We address this gap using colored inks as distinguishable information carriers. The key idea is to inject cyan (C), magenta (M), and yellow (Y) inks into a shared background-flow channel and estimate the resulting color traces through non-invasive spectral sensing. This creates three simultaneous molecule-domain subchannels from one optical measurement and supports both \ac{MUMO}-\ac{OOK} transmission and future molecule-domain multiple-access experiments. The same physical principle has also been used as a low-cost educational testbed for hands-on \ac{MC} experiments~\cite{gaedekenHandsOnMolecularCommunication2025}, which highlights the accessibility of color-based spectral sensing beyond this study.

This paper builds on our conference version in~\cite{wietfeldEvaluationMultiMoleculeMolecular2024b}, which introduced the spectral \ac{MUMO} testbed, the C/M/Y spectral unmixing pipeline, and initial proof-of-concept real-time \ac{MUMO}-\ac{OOK} measurements. The present journal paper extends this starting point through longer-distance isolated-pulse measurements, effective end-to-end modeling, a broad detector-family benchmark on continuous payloads, and practical lessons from repeated testbed operation.
The main contributions are as follows:
\begin{enumerate}
    \item We extend the real-time spectral \ac{MUMO} testbed through refined fluidic, mechanical, and electrical implementation, and document its construction and operation through a detailed component inventory and empirical practical lessons.
    \item We characterize isolated-pulse measurements as effective end-to-end responses shaped by finite release, flow propagation, \ac{RX} smoothing, amplitude scaling, and timing alignment.
    \item We compare an \ac{AD}-style effective broadening model and an inner-disk Poiseuille benchmark under a common fitting policy, using one global broadening or radial-support parameter across all modeled colors and distances.
    \item We compare trace-only and \ac{CIR}-assisted detector families on \(6600\) measured simultaneous C/M/Y payload bits under a common retrospective full-data optimization protocol.
    \item We introduce \ac{MEDD}, which scales the previous-energy reference and adds a trace-adaptive offset to \ac{EDD}. It outperforms all other trace-only detectors with \(34/4500=0.76\%\) \ac{BER} and differs by only two errors from the best \ac{MMSE} benchmark.
\end{enumerate}

The remainder of this paper is structured as follows. Sec.~\ref{sec:related_work} reviews related work on experimental platforms, modeling, detectors, and \ac{MUMO} signaling. Sec.~\ref{sec:spectral_mumo_testbed} describes the spectral \ac{MUMO} testbed and measurement campaigns. Sec.~\ref{sec:effective_channel_modeling} presents the effective end-to-end modeling approach. Sec.~\ref{sec:detector_benchmark} evaluates detector families on measured payload traces, and Sec.~\ref{sec:conclusion} concludes the paper.

The supplementary material provides a detailed component inventory and links to the printable 3D models used for the testbed assembly. It further contains model derivations and parameter grids, detector definitions and search settings, and additional measurement data and results supporting the main analysis.

\section{Related Work}
\label{sec:related_work}

Experimental \ac{MC} research spans different molecule classes, propagation environments, \ac{RX} principles, and networking abstractions. Recent surveys summarize this growing experimental landscape~\cite{farsad_comprehensive_2016,lotterExperimentalResearchSynthetic2023e,lotterExperimentalResearchSynthetic2023f}. Most individual studies, however, still emphasize one layer of the communication chain, such as platform demonstration, channel modeling, \ac{RX} design, or networking concepts. In contrast, this work uses one real-time spectral \ac{MUMO} flow testbed as a common basis for effective end-to-end modeling and practical detector benchmarking. Multi-user and multiple-access concepts motivate the platform design, but the present paper focuses on experimentally supported \ac{MUMO} modeling and detection.

\subsection{Experimental Flow-Based \texorpdfstring{\acs{MC}}{MC} Testbeds}
\label{subsec:rw_testbeds}

Early and established flow-based \ac{MC} testbeds demonstrated practical liquid-channel communication using acid/base signaling, magnetic nanoparticles, and salinity changes~\cite{farsadNovelExperimentalPlatform2017,unterwegerExperimentalMolecularCommunication2018,wickeExperimentalSystemMolecular2022,angerbauerSalinityBasedMolecularCommunication2023b}. These platforms demonstrate physical \ac{MC} links in controlled fluidic environments, but they mostly focus on one signaling mechanism and one \ac{RX} pipeline. More application-oriented and biologically inspired platforms have studied biocompatible magnetic nanoparticles, reusable measurement data for channel-parameter studies, protein switching, chemical-domain signal processing, and in vivo vascular environments~\cite{bartunikDevelopmentBiocompatibleTestbed2023,bartunikChannelParameterStudies2023,schererClosedLoopLongTermExperimental2026,walterRealtimeSignalProcessing2023,vakilipoorCAMModelVivo2025}. These systems increase realism or chemical sophistication, but they do not generally target low-cost reconfigurability together with communication-metric-oriented measurement campaigns.

Color-based and \ac{MUMO} experimental systems have shown that distinguishable molecular species can support richer signaling than single-molecule links~\cite{panMolecularCommunicationPlatform2022,caliExperimentalImplementationMolecule2024b,bartunikColourspecificMicrofluidicDroplet2020}. However, many of these systems either do not transmit multiple molecule types simultaneously in real time, or they do not use separated traces for systematic detector benchmarking. Our conference precursor introduced a low-cost spectral \ac{MUMO} testbed with C/M/Y ink, non-invasive spectral sensing, linear intensity estimation, initial \ac{CIR} measurements, and proof-of-concept \ac{MUMO}-\ac{OOK} communication~\cite{wietfeldEvaluationMultiMoleculeMolecular2024b}. The present paper extends this from a platform demonstration toward distance-dependent effective modeling and a broader detector benchmark.

\subsection{Channel and Injection Modeling for Flow-Based \texorpdfstring{\acs{MC}}{MC}}
\label{subsec:rw_channel_injection}

General \ac{MC} channel-modeling work provides the analytical basis for diffusion, advection, reactions, \ac{RX} effects, release mechanisms, and end-to-end \acp{CIR}~\cite{jamaliChannelModelingDiffusive2019}. This supports the effective end-to-end viewpoint used here, where the measured response includes \ac{TX} release, physical transport, and \ac{RX} smoothing rather than pure propagation alone. Pipe-flow models describe laminar advection, diffusion, and effective dispersion in cylindrical channels~\cite{schaferTransferFunctionModels2021,wickeExperimentalSystemMolecular2022}. These models are essential references, but practical injection systems can violate idealized assumptions such as instantaneous release, known initial cross-sectional distribution, and perfectly separable propagation dynamics.

Experimental studies of liquid-channel complexity and fluid-dependent propagation show that real \ac{MC} measurements can deviate substantially from simplified models~\cite{wangUnderstandingEmbracingComplexities2020b,debusBloodMakesDifference2025}. This motivates using measurement-driven effective models and interpreting fitted parameters cautiously. Recent work on injection design and reduced hemodynamic transport models further shows that release geometry and flow-dominated residence-time effects can strongly influence the measured signal~\cite{willeEffectInjectionDesign2025,jakumeitMixtureInverseGaussians2025}. Across these bodies of work, a clear missing intersection remains: practical flow-based \ac{MUMO} testbeds still lack compact measurement-driven procedures that interpret finite release, flow propagation, effective broadening, and \ac{RX} effects.

\subsection{Detection and Receiver Processing}
\label{subsec:rw_detection}

Prior work on \ac{MC} \ac{RX} design has studied threshold, energy, sequence, equalization, and model-based detectors under idealized or semi-analytical signal models~\cite{kilincReceiverDesignMolecular2013,llatserDetectionTechniquesDiffusionbased2013,mengReceiverDesignDiffusionBased2014,qianMolecularCommunicationsModelBased2019a}. These detector classes are valuable baselines, but their assumptions are often difficult to satisfy in real flow measurements with timing drift, baseline variation, and \ac{CIR} mismatch. Low-complexity adaptive and noncoherent detectors address the practical difficulty of obtaining stable channel knowledge~\cite{damrathLowComplexityAdaptiveThreshold2016,alshammriLowcomplexityMemoryassistedAdaptivethreshold2017,noelAsynchronousPeakDetection2017}. The \ac{DD} and \ac{EDD} baselines, together with the \ac{MEDD} variant introduced in this paper, follow this practical direction by using trace-derived reference differences instead of explicit \ac{CIR}-assisted \ac{ISI} cancellation.

\ac{CIR}-assisted \acp{MF}, \ac{MMSE} equalizers, and sequence detectors provide principled model-assisted \ac{RX} classes when templates or channel estimates are available~\cite{jamaliDesignMatchedFilters2017,kilincReceiverDesignMolecular2013,mengReceiverDesignDiffusionBased2014,qianMolecularCommunicationsModelBased2019a}. In experimental testbeds, their performance depends on how representative these templates remain across colors, payloads, timing offsets, and operating points. Data-driven and testbed-optimized \ac{RX}s have shown that learning-based or measurement-tuned methods can work when analytical models are unreliable~\cite{farsadNovelExperimentalPlatform2017,hofmannTestbedbasedReceiverOptimization2022,heinleinClosingImplementationGap2024}. However, limited measurement volume and retraining burden can make them difficult to apply to small experimental \ac{MC} datasets. This motivates the detector benchmark in Sec.~\ref{sec:detector_benchmark}.

\subsection{Multi-Molecule and Multiple-Access \texorpdfstring{\acs{MC}}{MC}}
\label{subsec:rw_mumo_ma}

\Ac{MUMO} signaling has been studied through color-based platforms, molecule shift keying, molecule-mixture signaling, and olfaction-inspired \ac{RX} models~\cite{panMolecularCommunicationPlatform2022,caliExperimentalImplementationMolecule2024b,jamaliOlfactioninspiredMCsMolecule2023}. These works show the value of distinguishable molecule types, but they do not generally combine real-time molecular signal separation, modeling, and detector benchmarking. Multi-user and \ac{MA} concepts have also been investigated through bacterial amplitude-division access, \ac{TDMA}-based drug release and data gathering, and resource allocation for medical \ac{MC} networks~\cite{krishnaswamyADMAAmplitudeDivisionMultiple2017a,rudsariDrugReleaseManagement2019,shitiriTDMABasedDataGathering2021,chenResourceAllocationMultiuser2021}. Practical multi-user molecular networks and salinity-based experimental platforms show that \ac{MC} can move beyond single-link demonstrations~\cite{wangPracticalScalableMolecular2023}. Prior work on \ac{NOMA} and chemical-domain successive interference cancellation motivates future non-orthogonal and reaction-based \ac{MA} experiments in \ac{MC}~\cite{wietfeldDBMCaNOMAlyAsynchronousNOMA2026,wietfeldDBMCNOMAEvaluatingNOMA2024,wietfeldErrorProbabilityOptimization2024c,wietfeldChemSICalEvaluatingStochastic2025}. 

The present paper addresses the experimental prerequisites for such work: real-time molecular signal separation, effective channel behavior, and detector choices that are benchmarked on measured simultaneous \ac{MUMO} payload traces.

\section{Spectral \texorpdfstring{\acs{MUMO}}{MUMO} Testbed}
\label{sec:spectral_mumo_testbed}

\begin{figure}[t]
    \centering
    \includegraphics[width=\linewidth]{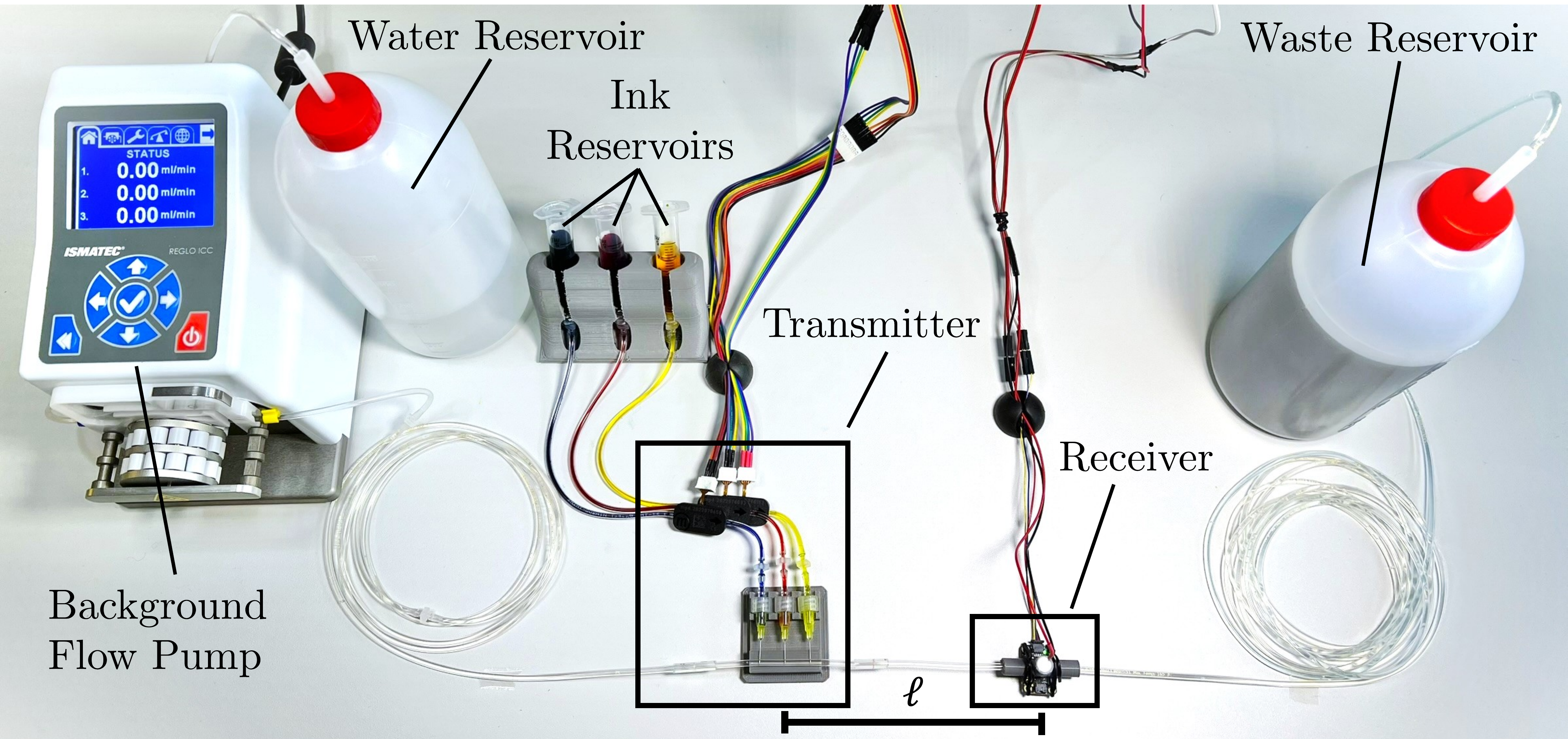}
    \caption{Annotated overview of the spectral \acs{MUMO} testbed. A peristaltic pump generates the background flow from a water reservoir through a transparent tube. The \acs{TX} injects cyan, magenta, and yellow inks through three micropump-driven needles. Downstream, the non-invasive \acs{RX} estimates the color intensities from an eight-channel spectral measurement before the mixture reaches the waste reservoir.}
    \label{fig:testbed_annotated}
\end{figure}

We use the real-time spectral \ac{MUMO} \ac{MC} testbed introduced in our conference precursor~\cite{wietfeldEvaluationMultiMoleculeMolecular2024b}. Here, we extend its use toward distance-dependent \ac{CIR} characterization and detector benchmarking. The setup consists of a liquid-flow channel, a three-color micropump \ac{TX}, and a non-invasive spectral \ac{RX}, as illustrated in Fig.~\ref{fig:testbed_annotated}.

The key design idea is to use distinguishable ink colors as parallel molecular subchannels. After spectral unmixing, the C/M/Y colors yield three estimated color-intensity traces that can be used both for channel modeling and for \ac{MUMO}-\ac{OOK} detector evaluation.

\subsection{Hardware and Flow Configuration}
\label{subsec:testbed_hardware_flow}

The channel is a transparent Tygon tube with diameter \(d_c=2r_c=\qty{1.6}{\milli\meter}\). The tube provides a simple reconfigurable flow channel while allowing optical access for the spectral \ac{RX}. Background flow is generated by a peristaltic pump~\cite{masterflexRegloICCBrochure}.

Compared to the conference setup~\cite{wietfeldEvaluationMultiMoleculeMolecular2024b}, the background-flow path includes an Innofluid fluid pulse damper for the \(\qty{1.6}{\milli\meter}\) inner-diameter tubing~\cite{innofluidPulseDamper2026}. The damper is used as a practical setup refinement to reduce peristaltic-pump pulsation effects in the background flow. The \ac{TX} and \ac{RX} interfaces are realized using custom 3D-printed housings, and the control electronics are mounted on a custom two-layer bare rigid \ac{PCB}. This modular structure allows the physical distance, needle arrangement, and \ac{RX} placement to be adapted across measurement campaigns more easily than in fully integrated microfluidic platforms.

\begin{figure}[t]
    \centering
    \includegraphics[width=\linewidth]{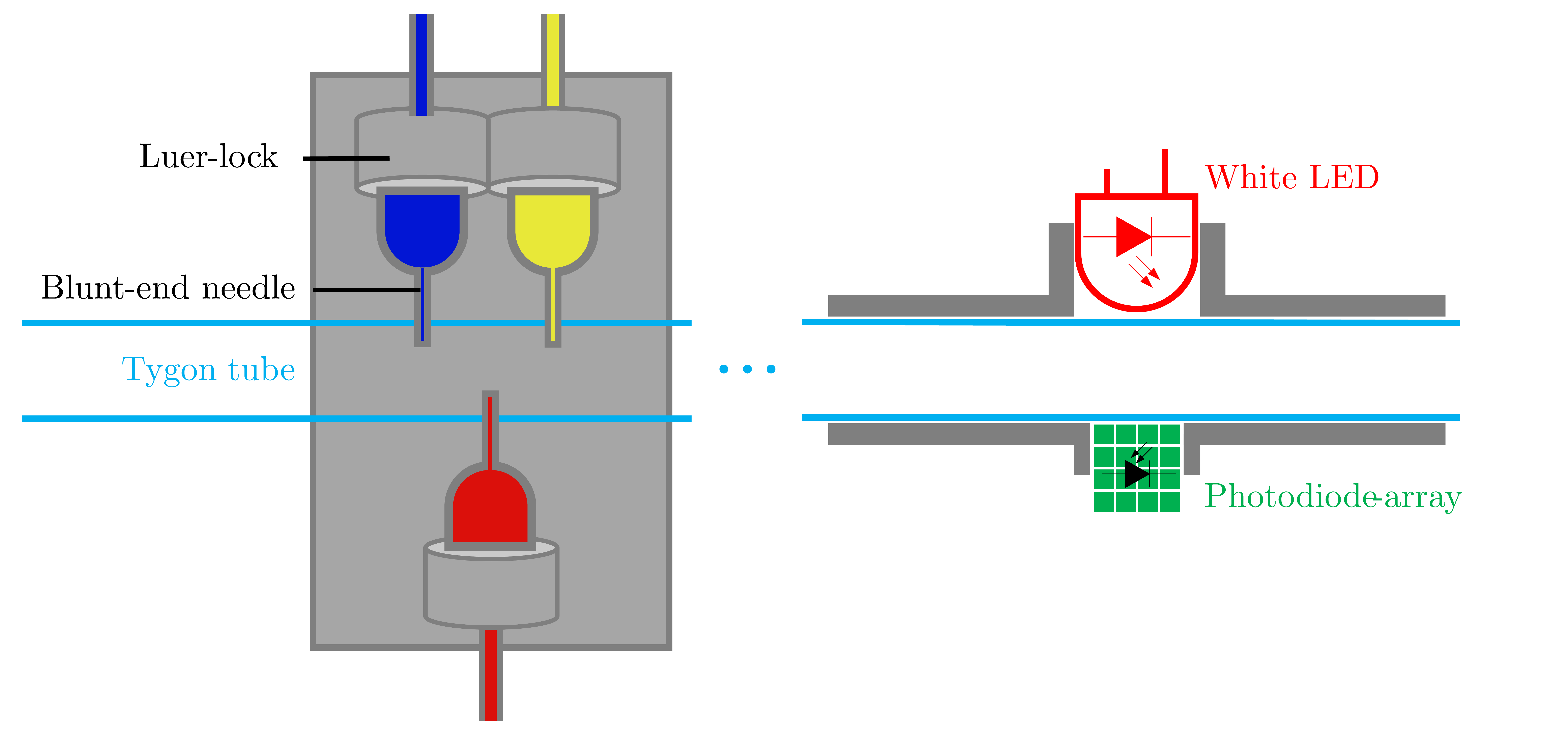}
    \caption{Schematic \acs{TX} and \acs{RX} geometry details. The \acs{TX} panel shows the luer-lock connector and blunt-end needle arrangement used for color injection into the background flow. The colored needles indicate the C/M/Y injection paths; cyan is upstream, magenta is centered, and yellow is downstream in the channel layout. The \acs{RX} panel shows the non-invasive optical readout geometry, where a white LED illuminates the tube and the AS7341 photodiode array records the transmitted spectral intensity.}
    \label{fig:tx_rx_closeups}
\end{figure}

The testbed should be interpreted as a mesoscopic flow-based \ac{MC} platform, not as a direct biomedical implementation. The use of printer inks provides inexpensive and well-separated optical signatures for the present experiments, while application-specific molecule types and sensing modalities would be needed for in-body scenarios.
Fig.~\ref{fig:tx_rx_closeups} complements the system-level view in Fig.~\ref{fig:testbed_annotated} by showing the schematic \ac{TX} needle/tube arrangement and the optical \ac{RX} readout geometry around the transparent channel.

\subsection{Transmitter and Information Molecules}
\label{subsec:testbed_tx_molecules}

The information molecules are cyan, magenta, and yellow printer inks, denoted by \(\mathcal C=\{\mathrm C,\mathrm M,\mathrm Y\}\). Their distinct absorption spectra allow the \ac{RX} to estimate the three color intensities from a mixed optical measurement.

The \ac{TX} consists of three Bartels BP7 piezoelectric micropumps~\cite{bartelsBp7TubingDatasheet2024}, each connected to one ink reservoir. The pumps are controlled through a driver circuit~\cite{bartelsElectronicDriverDatasheet2025} and an Arduino Micro. The ink is injected through a blunt-end needle with outer diameter \(\qty{0.23}{\milli\meter}\) and inner diameter \(\qty{0.11}{\milli\meter}\). The needles are aligned approximately perpendicular to the flow direction and separated by \(\qty{1.3}{\centi\meter}\) along the channel.

The needle ordering is cyan upstream, magenta in the middle, and yellow downstream. The needle insertion depths are adjusted to reduce pulse distortion as released ink passes the other needles, but the exact arrangement remains a practical source of color-dependent release differences. Binary \ac{OOK} is implemented independently for each color \(c\in\mathcal C\). For symbol \(k\), bit \(b_{k,c}=1\) triggers a pump voltage \(V_1=\qty{110}{\volt}\) for an injection duration \(T_{\mathrm{inj}}\). Bit \(b_{k,c}=0\) uses the idle level \(V_0=V_{\mathrm{idle}}=\qty{40}{\volt}\). The voltages cannot be mapped to exact ink volumes due to variations among injection events.

The idle voltage is applied continuously to mitigate backflow into the \ac{TX} needles caused by the pressure difference between the micropump outlet and the background flow. This makes \(V_{\mathrm{idle}}\) a hardware-stability parameter corresponding to a communication symbol amplitude of zero. During \ac{MUMO}-\ac{OOK} transmission, the three colors are driven in parallel, so one aggregate symbol can contain up to three simultaneously transmitted bits. For a symbol duration \(T_{\mathrm{sym}}\), the nominal aggregate bit rate is \(R_{\mathrm{agg}}=3/T_{\mathrm{sym}}\) when all three color subchannels are active.

\subsection{Spectral Receiver and Color-Intensity Estimation}
\label{subsec:testbed_rx_spectral_estimation}

The \ac{RX} uses an Adafruit AS7341 spectral sensor with eight visible channels centered at \(415\), \(445\), \(480\), \(515\), \(555\), \(590\), \(630\), and \(\qty{680}{\nano\meter}\)~\cite{amsOSRAMAS7341Datasheet2020}. The sensor is placed in a custom 3D-printed housing together with a white LED that illuminates the tube from the opposite side. The sensor and LED are controlled by an Arduino Micro, and the \ac{RX} samples the spectral measurements at approximately \(\qty{19}{\hertz}\). The \ac{RX} is non-invasive because it observes light attenuation through the tube without inserting an electrode, probe, or particle sensor into the flow. Optical attenuation through a transparent tube is not itself an in-body \ac{RX} architecture. Here it is used as a simple stand-in for non-invasive sensing, while biomedical use would require compatible selective sensing modalities.

Let \(I^j(t)\) be the measured light intensity in spectral channel \(j\), and let \(I_0^j\) be the corresponding no-ink reference intensity. We model the absorbance in channel \(j\) using the Beer-Lambert law~\cite{swinehartBeerLambertLaw1962} as
\begin{equation}
    a^j(t)
    =
    \log_{10}\!\left(\frac{I_0^j}{I^j(t)}\right).
    \label{eq:absorbance}
\end{equation}
For color \(c\in\mathcal C\), the color-intensity trace \(\chi_c(t)\) is related to the absorbance contribution in channel \(j\) through
\begin{equation}
    a_c^j(t)=b_c^j \chi_c(t),
    \label{eq:beer_lambert_matrix_element}
\end{equation}
where \(b_c^j\) is the absorption coefficient of color \(c\) in spectral channel \(j\). Assuming additive absorbance contributions from the three inks~\cite{griffithsFourierTransformInfrared2007}, the measured absorbance vector can be written as
\begin{equation}
    \mathbf a(t)
    =
    \mathbf B \boldsymbol{\chi}(t),
    \qquad
    \boldsymbol{\chi}(t)
    =
    \begin{bmatrix}
        \chi_{\mathrm C}(t) &
        \chi_{\mathrm M}(t) &
        \chi_{\mathrm Y}(t)
    \end{bmatrix}^{\mathsf T},
    \label{eq:absorbance_vector_model}
\end{equation}
where \(\mathbf a(t)\in\mathbb R^8\) and \(\mathbf B=[\mathbf b_{\mathrm C}\ \mathbf b_{\mathrm M}\ \mathbf b_{\mathrm Y}]\in\mathbb R^{8\times 3}\).

Before each measurement, a single-color calibration is conducted to obtain the absorption matrix \(\mathbf B\). For color \(c\), the corresponding column is estimated by averaging the peak absorbance vectors over the calibration-pulse intervals \(\mathcal K_c\),
\begin{equation}
    \mathbf b_c
    =
    \operatorname{mean}_{\mathcal T\in\mathcal K_c}
    \max_{t\in\mathcal T}
    \mathbf a(t),
    \label{eq:calibration_matrix}
\end{equation}
where each \(\mathcal T\) is the reception interval of one calibration pulse, and the maximum over \(\mathbf a(t)\) is taken componentwise over the spectral channels. The estimated color-intensity vector is obtained by ordinary least squares using the Moore-Penrose pseudoinverse~\cite{kayFundamentalsStatisticalSignal1993},
\begin{equation}
    \hat{\boldsymbol{\chi}}(t)
    =
    \mathbf B^\dagger \mathbf a(t)
    =
    \left(\mathbf B^{\mathsf T}\mathbf B\right)^{-1}
    \mathbf B^{\mathsf T}\mathbf a(t),
    \label{eq:spectral_unmixing}
\end{equation}
where the last equality is the full-column-rank least-squares form used for the calibrated absorption matrix in the considered measurements.

\begin{figure}[t]
    \centering
    \includegraphics[width=\linewidth]{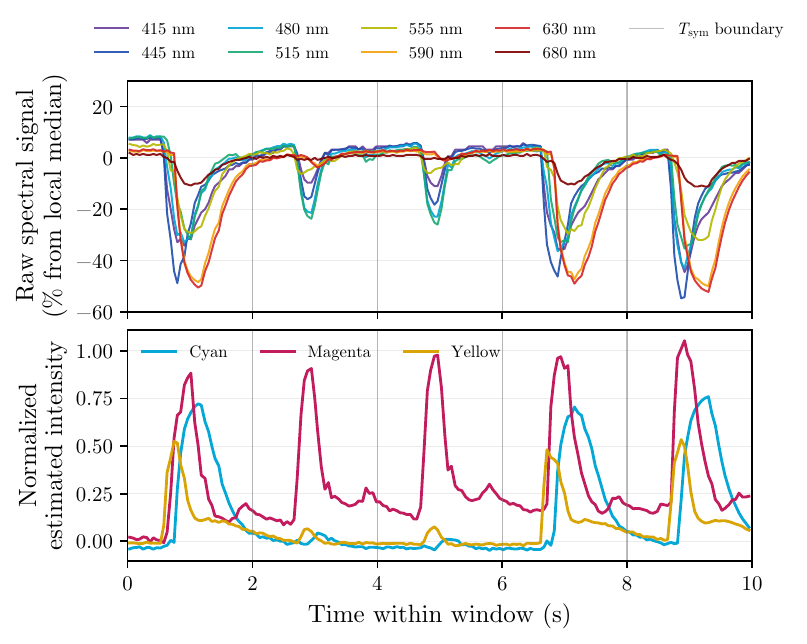}
    \caption{Raw spectral \acs{RX} channels and converted color-intensity estimates for a representative \(\qty{10}{\second}\) section of the short-channel run with symbol period \(T_\mathrm{sym}=\qty{2}{\second}\) and injection duration \(T_\mathrm{inj}=\qty{0.2}{\second}\). The raw channels are shown as deviations from the median only for display; estimation uses the calibrated intensities in \eqref{eq:spectral_unmixing}.}
    \label{fig:example_raw_estimated_trace}
\end{figure}

The resulting traces \(\hat\chi_{\mathrm C}(t)\), \(\hat\chi_{\mathrm M}(t)\), and \(\hat\chi_{\mathrm Y}(t)\) are treated as the observable \ac{RX} signals throughout the paper. They are dimensionless color-intensity estimates and should not be interpreted as absolute molecule counts or absolute molar concentrations. For compact notation in the detector section, we denote the estimated trace of color \(q\) by \(y_q(t)=\hat\chi_q(t)\). For the modeling section, the same traces are used to construct measured \acp{CIR} from isolated-pulse measurements. Fig.~\ref{fig:example_raw_estimated_trace} shows a representative raw-to-estimated trace segment and the observable signal format used in the modeling and detector sections, Secs.~\ref{sec:effective_channel_modeling} and~\ref{sec:detector_benchmark}.

\subsection{Measurement Campaigns and Datasets}
\label{subsec:testbed_measurement_campaigns}

Two types of measurements are used in the paper: (i) isolated-pulse measurements for \ac{CIR} construction and effective end-to-end modeling, and (ii) continuous random payload measurements for detector benchmarking. The main modeling campaign uses isolated C/M/Y pulse measurements at \(\ell=\qty{9.5}{\centi\meter}\), \(\qty{23.5}{\centi\meter}\), and \(\qty{100}{\centi\meter}\). These measurements are analyzed as effective \acp{CIR} after pulse extraction and alignment. 
The detector benchmark uses continuous \ac{MUMO}-\ac{OOK} payload measurements at \(\ell=\qty{7.4}{\centi\meter}\) and \(\ell=\qty{23.5}{\centi\meter}\). The evaluated operating points vary \(T_{\mathrm{sym}}\) and \(T_{\mathrm{inj}}\) to study throughput, \ac{ISI}, injection-duration effects, and stress cases.
The \(\qty{7.4}{\centi\meter}\) detector channel and the \(\qty{9.5}{\centi\meter}\) modeling channel are called short-distance cases. The \(\qty{23.5}{\centi\meter}\) channel is called the medium-distance case, and the \(\qty{100}{\centi\meter}\) channel is called the long-distance case.

\subsection{Practical Lessons Learned}
\label{subsec:practical_lessons}

The following lessons summarize empirical experience from repeated assembly and operation of the testbed. They are qualitative observations rather than controlled component comparisons. Besides supporting reproducible operation, they motivate several modeling and detector choices introduced in Secs.~\ref{sec:effective_channel_modeling} and~\ref{sec:detector_benchmark}.

\paragraph{Injection geometry}
Needle placement proved to be one of the most sensitive aspects of the setup. Small changes in insertion position, depth, or angle altered the effective release behavior and could change the complete received pulse shape. We reduced this variability using a 3D-printed \ac{TX} housing that fixed the injection structures during operation and a removable alignment collar that helped with needle centering, insertion depth, and uniform spacing during assembly. Links to the printable \ac{TX} and \ac{RX} housings, tube-hole insertion helper, and syringe holder are provided in the supplementary material. The thinnest practical needles were selected because larger openings increased both passive ink leakage and water intrusion. For the investigated setup, approximately perpendicular insertion provided the most stable empirical compromise: an opening directed against the background flow experienced strong water intrusion, whereas alignment with the flow increased passive ink release. Nevertheless, some color- and campaign-dependent release variability remained. This observation motivates the effective finite-release source model and case-wise source-shape parameters introduced in Sec.~\ref{subsec:modeling_source_release}, rather than assuming one ideal injection pulse for all measurements.

\paragraph{Pressure balance and dilution}
The injection paths remain part of the hydraulic system even when no active transmission is intended. Changes in the liquid levels of the background-water, ink, and waste reservoirs modified the pressure balance and could either push water into the ink paths or cause continuous ink leakage into the channel. During storage and longer periods of inactivity, plastic clamps placed around the tubes directly behind the needles isolated the ink paths and limited both effects.

Backflow remained relevant during operation because the active background flow continuously pressed water toward the needle openings. Before longer measurements, we therefore applied a several-second injection from all colors to replace diluted liquid inside the needles and connecting tubes. During transmission, the positive idle voltage \(V_{\mathrm{idle}}\) maintained a counterpressure during zero symbols. It was selected as the highest setting that did not produce visible leakage at the \ac{RX}. Even with this measure, gradual dilution could occur during long payload sequences and appeared as slowly decreasing pulse amplitudes. This motivates treating the measured amplitudes as operating-point-dependent observations rather than as known injected volumes in both the modeling and detector analyses.

\paragraph{Air bubbles and flow stability}
Air bubbles disturbed the pressure conditions inside the injection paths and produced large transient responses when passing through the optical \ac{RX}. For storage, the needles were therefore removed from the active channel and inserted into a separate sealed dummy channel while the main flow path was closed. This isolated the ink paths from the background-flow pressure and allowed bubbles to leave the elevated injection structures more easily. Measurements were started only after the reservoirs, tubing, and needles had been inspected and operated until no visible bubbles remained.

The peristaltic background pump introduced another directly observable effect. Without damping, an ink front inside the tube visibly moved forward and backward with the pump cycle. After adding the pulse damper described in Sec.~\ref{subsec:testbed_hardware_flow}, the visible flow became substantially smoother. Remaining flow irregularities and unmodeled fluidic effects can contribute to the effective-model residuals discussed in Sec.~\ref{subsec:modeling_results_limits}.

\paragraph{Mechanical, electrical, and sensing reliability}
The initial jumper-wire connections occasionally loosened without an obvious warning and caused intermittent pump or sensor failures. Replacing them with a soldered \ac{PCB}, fixed cables, and proper connectors substantially improved reliability. The custom 3D-printed \ac{TX}, \ac{RX}, and reservoir housings similarly reduced unintended component movement between measurements.

The Arduino-controlled transmission schedule also introduced a small processing delay that accumulated over long symbol sequences. We incorporate this effect as the tunable per-symbol timing correction \(\Delta\) in the detector timing model of Sec.~\ref{subsec:detector_protocol}. In addition, some spectral sensors exhibited reduced sensitivity after extended operation and were replaced. Prolonged strong illumination may have contributed, but the degradation mechanism was not independently identified. Reference and single-color responses were therefore checked before longer campaigns.

Finally, distilled or otherwise low-mineral water should be used throughout the setup. Tap water left mineral deposits after partial drying, which impaired optical measurements and could obstruct narrow tubing, pumps, and needles. Tubing and needles should therefore be treated as consumable components, and the fluidic paths should be cleaned before water or ink residues dry.


Not all distances and operating points originate from one perfectly identical hardware state. Therefore, campaign changes, needle placement, and timing offsets are treated as practical experimental factors that can cause noise-like variations and deviations from the expected behavior. These effects are reflected in the modeling diagnostics in Sec.~\ref{subsec:modeling_results_limits} and the detector \textit{Stress Settings} in Sec.~\ref{subsec:detector_stress_settings}.


\section{Effective End-to-End Modeling Across Distances}
\label{sec:effective_channel_modeling}

The isolated-pulse measurements are modeled as effective end-to-end responses rather than pure propagation kernels. Thus, release, propagation, and reception jointly shape the observed \ac{CIR}~\cite{jamaliChannelModelingDiffusive2019}. The modeling data set contains cyan, magenta, and yellow \acp{CIR} at \(\ell\in\{\qty{9.5}{\centi\meter},\qty{23.5}{\centi\meter},\qty{100}{\centi\meter}\}\), denoted as \textit{short-}, \textit{medium-}, and \textit{long-}channel cases. The objective is not to identify microscopic molecule parameters, but to obtain a compact and interpretable effective model for the measured pulse shapes across distance and color.

We use two candidate propagation/residence components under the same source, \ac{RX}, scaling, timing, and fitting policy. The \ac{AD}-style model serves as a compact mechanistic fitting model with nominal advection and effective longitudinal broadening. The inner-disk Poiseuille model serves as a simple laminar-flow-based comparison benchmark that captures the broadening effect via the parabolic flow speed variations.

\subsection{Measurement Set and Flow-Scale Timing}
\label{subsec:modeling_measurements_timing}

Let \(\ell\) denote the \ac{TX}--\ac{RX} distance. For compact notation, we index one color-distance \ac{CIR} case by \(j=(c,\ell)\). The measured \ac{CIR} for case \(j\) is denoted by \(y_j(\tau)\), where \(\tau\) is the local time coordinate used for the extracted isolated-pulse response, and the corresponding model prediction is denoted by \(\hat y_j(\tau)\).

The nominal mean flow speed is derived from the pump setting and channel radius as
\begin{equation}
    v_{\mathrm{avg}}
    =
    \frac{Q_0}{\pi r_c^2},
    \qquad
    Q_0=\qty{10}{\milli\liter\per\minute},
    \qquad
    r_c=\qty{0.8}{\milli\meter},
    \label{eq:e2e_mean_flow_speed}
\end{equation}
which gives \(v_{\mathrm{avg}}\approx\qty{0.0829}{\meter\per\second}\). We use this value as a nominal flow-scale reference, not as a fitted velocity measurement. Case-wise timing differences are represented by the alignment parameter introduced in the following subsection.

\subsection{End-to-End Effective Model}
\label{subsec:modeling_effective_model}

We model the measured response as a convolution of an effective finite-release source \(i_j(t)\), a propagation kernel \(h_j(t)\), and an effective \ac{RX} smoothing kernel \(r_{\mathrm{RX}}(t)\). The fitted model is
\begin{equation}
    \hat y_j(\tau)
    =
    A_j
    \left[
        i_j * h_j * r_{\mathrm{RX}}
    \right](\tau-s_j),
    \label{eq:e2e_observation}
\end{equation}
where \(*\) denotes convolution, \(A_j\ge 0\) is a nuisance amplitude scale, and \(s_j\) is a case-wise timing-alignment parameter. The residual modeling error is
\begin{equation}
    e_j(\tau)=y_j(\tau)-\hat y_j(\tau).
    \label{eq:e2e_residual}
\end{equation}
The residual contains unmodeled \ac{TX} behavior, remaining flow irregularities, \ac{RX} and spectral-estimation effects, and measurement noise.

All component kernels are normalized on the sampled model grid before convolution. With model time step \(\Delta t=\qty{0.01}{\second}\), the normalization operator is written in continuous form as
\begin{equation}
    \mathcal N_t[f](t)
    =
    \frac{f(t)}
    {\int_0^\infty f(\xi)\,\mathrm d\xi},
    \label{eq:e2e_numeric_normalization}
\end{equation}
and implemented by the corresponding finite-grid sum whenever the finite-grid area is positive and finite. The final convolved waveform in \eqref{eq:e2e_observation} is not renormalized before fitting. Therefore, \(A_j\) absorbs color-dependent intensity scale, injection amount, optical gain, and other amplitude effects, and should not be interpreted as an absolute molecule count or injected mass.

\begin{figure}[t]
    \centering
    \includegraphics[width=0.8\linewidth]{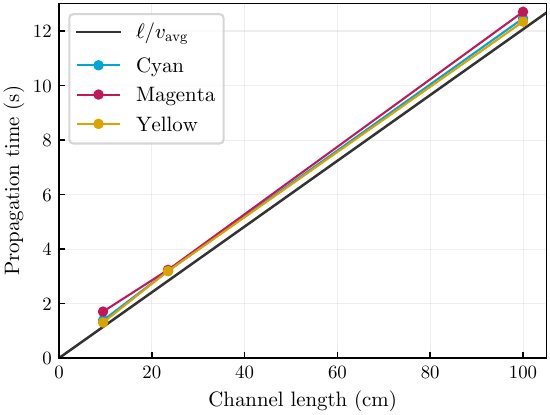}
    \caption{Flow-scale timing diagnostic for the modeled \acp{CIR}. Colored markers show \(-s_j\), the propagation-time quantity implied by the fitted timing-alignment parameter of the end-to-end fit. The solid line shows the nominal flow-scale time \(\ell/v_{\mathrm{avg}}\).}
    \label{fig:flow_time_scaling}
\end{figure}

Fig.~\ref{fig:flow_time_scaling} compares \(-s_j\), the propagation-time quantity implied by the case-wise fitted alignment shift, with the nominal flow-scale time \(\ell/v_{\mathrm{avg}}\). Their proximity supports a flow-scale interpretation of the dominant arrival timing.

\subsection{Finite-Release Source Characteristics}
\label{subsec:modeling_source_release}

The micropump and needle \ac{TX} does not produce an ideal impulse release. We therefore model the \ac{TX} contribution through an effective finite-release source profile \(i_j(t)\), which is fitted jointly with the propagation/residence and \ac{RX} components. Let \(T_j\) denote the pulse duration for case \(j\). The command \(c_{T_j}(t)\) is the idealized pump-control input: a unit-amplitude pulse that is active during the intended injection interval and zero otherwise, with cosine-smoothed edges to avoid an artificial discontinuity. The exact edge-duration convention is listed in the supplement.

The finite-release source is constructed from two area-normalized components. The first component is the commanded injection pulse \(c_{T_j}(t)\), and the second is a smoothed release state \(x_j(t)\). We write
\begin{equation}
    \bar c_{T_j}(t)=\mathcal N_t[c_{T_j}](t),
    \qquad
    \bar x_j(t)=\mathcal N_t[x_j](t),
    \label{eq:e2e_source_components}
\end{equation}
where \(\mathcal N_t[\cdot]\) denotes the time-area normalization in \eqref{eq:e2e_numeric_normalization}. The raw blended source shape is
\begin{equation}
    \tilde i_j(t)
    =
    (1-\lambda_j)\bar c_{T_j}(t)
    +
    \lambda_j\bar x_j(t),
    \label{eq:e2e_source_profile_raw}
\end{equation}
with mixture weight \(\lambda_j\in[0,1]\). Thus, \(\lambda_j=0\) gives the normalized command pulse, whereas \(\lambda_j=1\) gives the normalized smoothed release state. The source kernel used in the end-to-end convolution is finally normalized as
\begin{equation}
    i_j(t)=\mathcal N_t[\tilde i_j](t).
    \label{eq:e2e_source_profile}
\end{equation}

The smoothed release state \(x_j(t)\) is modeled as a first-order dynamic response to the smoothed command pulse, with separate effective build and decay time constants. Low-order dynamic states are standard tools for representing actuator behavior with memory in system identification~\cite{ljung1999system}. The asymmetric switching between build and decay constants is mathematically analogous to attack--release one-pole smoothing~\cite{giannoulisDigitalDynamicRange2012}, while the inclusion of a finite-duration source component follows the end-to-end \ac{MC} distinction between transmitter release and subsequent propagation~\cite{jamaliChannelModelingDiffusive2019}. On the model grid \(t_n\), the state is initialized as \(x_{j,0}=0\) and evaluated recursively as
\begin{equation}
    x_{j,n}
    =
    \max\left\{
        0,\,
        x_{j,n-1}
        +
        \Delta t
        \frac{c_{T_j}(t_n)-x_{j,n-1}}{\tau_{j,n}}
    \right\},
    \label{eq:e2e_source_state}
\end{equation}
with
\begin{equation}
    \tau_{j,n}
    =
    \begin{cases}
        \tau_{\mathrm{build},j}, & c_{T_j}(t_n)>x_{j,n-1},\\
        \tau_{\mathrm{decay},j}, & \text{otherwise}.
    \end{cases}
    \label{eq:e2e_source_state_tau}
\end{equation}
At each step, the current command sample determines whether the release state rises toward the command using \(\tau_{\mathrm{build},j}\) or relaxes using \(\tau_{\mathrm{decay},j}\). Thus, the release waveform depends on its previous value, not only on the instantaneous command sample.

The parameters \(\lambda_j\), \(\tau_{\mathrm{build},j}\), and \(\tau_{\mathrm{decay},j}\) are effective source-shape parameters. They compactly describe observed release broadening and should not be interpreted as direct measurements of pump leakage, pressure relaxation, molecular trapping, or emitted volume.

\begin{figure}[t]
    \centering
    \includegraphics[width=0.8\linewidth]{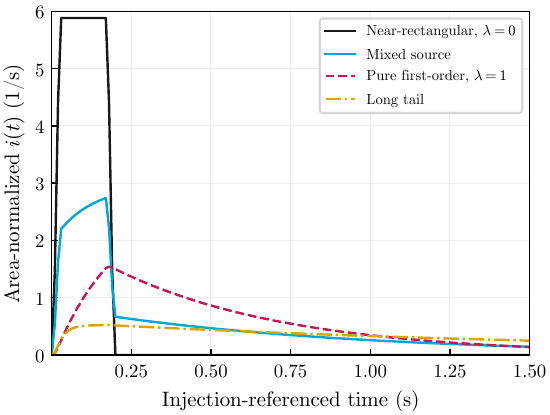}
    \caption{Schematic finite-release source-profile examples. The curves are area-normalized synthetic profiles that illustrate near-rectangular, mixed, first-order, and long-tail release behavior in the source model before propagation and \acs{RX} smoothing are applied.}
    \label{fig:finite_release_examples}
\end{figure}

Fig.~\ref{fig:finite_release_examples} gives a compact visual guide to the finite-release source family by showing normalized example profiles ranging from near-rectangular release to long-tail release. These schematic curves are not fitted data, but they clarify how the source model changes before it is convolved with the propagation/residence and \ac{RX} components.

For intuition, the source model contains several useful limiting cases. In the following, we use the idealized rectangular command \(\tilde c_{T_j}(t)\) with the case-specific pulse duration \(T_j\),
\begin{equation}
    \tilde c_{T_j}(t)
    =
    \begin{cases}
        1, & 0\le t\le T_j,\\
        0, & \text{otherwise},
    \end{cases}
\end{equation}
instead of the cosine-smoothed implementation used in the numerical fit.

\paragraph{Near-rectangular release}
If \(\lambda_j=0\), the smoothed release state is ignored and the source becomes the normalized command pulse,
\begin{equation}
    i_j(t)
    =
    \begin{cases}
        1/T_j, & 0\le t\le T_j,\\
        0, & \text{otherwise}.
    \end{cases}
    \label{eq:source_edge_rect_lambda}
\end{equation}
The same limiting behavior is obtained for any \(\lambda_j\) if the release state follows the command instantaneously,
\begin{equation}
    \tau_{\mathrm{build},j},\tau_{\mathrm{decay},j}\rightarrow 0
    \ \Rightarrow\ 
    x_j(t)\rightarrow \tilde c_{T_j}(t),
    \ 
    i_j(t)\rightarrow \mathcal N_t[\tilde c_{T_j}](t).
    \label{eq:source_edge_rect_tau}
\end{equation}

\paragraph{Pure first-order finite release}
If \(\lambda_j=1\), the source is fully determined by the smoothed release state,
\begin{equation}
    i_j(t)=\mathcal N_t[x_j](t).
\end{equation}
For the rectangular command and continuous-time limit, this state is
\begin{equation}
    x_j(t)
    =
    \begin{cases}
        1-\mathrm e^{-\frac{t}{\tau_{\mathrm{build},j}}},
        & 0\le t\le T_j,\\[0.5ex]
        \left(1-\mathrm e^{-\frac{T_j}{\tau_{\mathrm{build},j}}}\right)
        \mathrm e^{-\frac{t-T_j}{\tau_{\mathrm{decay},j}}},
        & t>T_j,\\[0.5ex]
        0,
        & t<0.
    \end{cases}
    \label{eq:source_edge_first_order}
\end{equation}
This corresponds to a release that rises with an effective build time and decays with an effective release-tail time.

\paragraph{Long-tail release}
If \(\lambda_j\approx1\), \(\tau_{\mathrm{build},j}\ll T_j\), and \(\tau_{\mathrm{decay},j}\gg T_j\), the state reaches approximately one during the commanded pulse and then decays slowly:
\begin{equation}
    x_j(t)
    \approx
    \begin{cases}
        1,
        & 0\le t\le T_j,\\[0.5ex]
        \mathrm e^{-\frac{t-T_j}{\tau_{\mathrm{decay},j}}},
        & t>T_j,\\[0.5ex]
        0,
        & t<0.
    \end{cases}
    \label{eq:source_edge_long_tail}
\end{equation}
Thus, \(i_j(t)=\mathcal N_t[x_j](t)\) represents an effective finite release with a residual tail. This behavior contrasts with the near-rectangular limiting case.

\begin{figure}[t]
    \centering
    \includegraphics[width=0.8\linewidth]{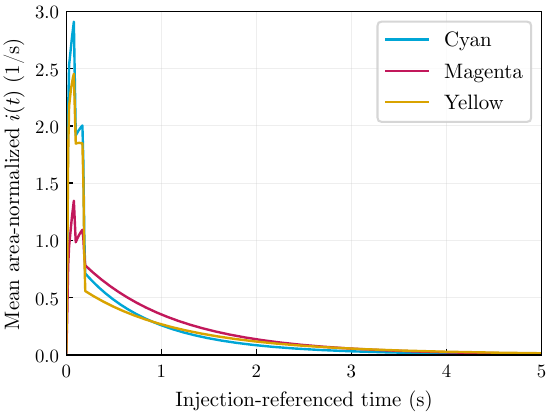}
    \caption{The plotted curves are arithmetic means of the area-normalized source profiles \(i_j(t)\) fitted across the selected distances. These profiles describe effective release broadening inside the end-to-end model and should not be interpreted as measured injection parameters.}
    \label{fig:injection_shapes}
\end{figure}

Fig.~\ref{fig:injection_shapes} shows the fitted source profiles averaged by color over the selected distances. The figure indicates that color-dependent finite release is a relevant part of the effective end-to-end response and differs visibly from the idealized rectangular pulse injection.

\subsection{Candidate Propagation and Residence Kernels}
\label{subsec:modeling_candidate_kernels}

We vary the propagation component \(h_j(t)\) in \eqref{eq:e2e_observation} between two compact models. Both use the same finite-release source, \ac{RX} smoothing, amplitude scaling, timing shift, and fitting objective. Differences between their fits can therefore be attributed to the propagation description rather than to a different fitting policy.

The \ac{AD}-style model uses a one-dimensional nominal-advection kernel with effective broadening. For \(t>0\), the raw kernel is
\begin{equation}
    g_{\mathrm{AD}}(t;\ell,B_{\mathrm{eff}})
    =
    \frac{1}{\sqrt{4\pi B_{\mathrm{eff}}t}}
    \exp\left(
        -\frac{(\ell-v_{\mathrm{avg}}t)^2}{4B_{\mathrm{eff}}t}
    \right),
    \label{eq:e2e_ad_raw}
\end{equation}
and \(g_{\mathrm{AD}}(t;\ell,B_{\mathrm{eff}})=0\) for \(t\le0\). The normalized \ac{AD}-style kernel is
\begin{equation}
    h_{\mathrm{AD}}(t;\ell,B_{\mathrm{eff}})
    =
    \mathcal N_t[g_{\mathrm{AD}}(t;\ell,B_{\mathrm{eff}})](t).
    \label{eq:e2e_ad_kernel}
\end{equation}
The scalar \(B_{\mathrm{eff}}\) is a global effective broadening coefficient of the fitted end-to-end model. Its unit follows from the kernel form and should not be read as a measured diffusion coefficient. It is not interpreted as a molecular diffusion coefficient, a turbulent diffusion coefficient, or a validation of a classical Taylor--Aris dispersion regime.

The inner-disk Poiseuille benchmark assumes a laminar velocity profile
\begin{equation}
    u(r)
    =
    u_0
    \left(
        1-\frac{r^2}{r_c^2}
    \right)
    =
    2v_{\mathrm{avg}}
    \left(
        1-\frac{r^2}{r_c^2}
    \right),
    \qquad 0\le r\le r_c,
    \label{eq:e2e_laminar_velocity_profile}
\end{equation}
with centerline speed $u_0=2v_{\mathrm{avg}}$.
The initial molecule support is represented by an idealized centered disk with radius \(r_i=\alpha r_c\), where \(0<\alpha\le1\), \(\alpha=r_i/r_c\), and \(s_i=\alpha^2\). This is used as a radial-support benchmark rather than as a literal side-needle plume model. For a finite \ac{RX} window of effective length \(\ell_{\mathrm{RX}}\), we define
\begin{equation}
    x_-=\ell-\frac{\ell_{\mathrm{RX}}}{2},
    \qquad
    x_+=\ell+\frac{\ell_{\mathrm{RX}}}{2},
    \qquad
    \ell_{\mathrm{RX}}=\qty{5}{\milli\meter}.
    \label{eq:e2e_disk_window}
\end{equation}
For \(t>0\), define the clipped window endpoints \(x_{\mathrm{up}}(t)=\min(x_+,u_0t)\) and \(x_{\mathrm{low}}(t)=\max(x_-,u_0t(1-s_i))\), and let \([x]_+=\max(0,x)\). The implemented raw inner-disk residence kernel is
\begin{equation}
    g_{\mathrm{disk}}(t;\ell,\alpha)
    =
    \frac{
        \left[
            x_{\mathrm{up}}(t)-x_{\mathrm{low}}(t)
        \right]_+
    }{s_i u_0t},
    \qquad t>0,
    \label{eq:e2e_disk_raw}
\end{equation}
and \(g_{\mathrm{disk}}(t;\ell,\alpha)=0\) for \(t\le0\). The normalized kernel is
\begin{equation}
    h_{\mathrm{disk}}(t;\ell,\alpha)
    =
    \mathcal N_t[g_{\mathrm{disk}}(t;\ell,\alpha)](t).
    \label{eq:e2e_disk_kernel}
\end{equation}
The disk kernel is included as a comparison because side-needle injection is neither an ideal point release nor a uniform release over the whole tube cross-section. The parameter \(\alpha=r_i/r_c\) is an effective radial-support parameter of this benchmark model, not a measured injection radius or plume radius. A compact derivation of \eqref{eq:e2e_disk_raw} from the Poiseuille velocity profile and the finite \ac{RX} window is provided in the supplementary material.

\subsection{Receiver Smoothing and Fitting Policy}
\label{subsec:modeling_fit_policy}

Receiver and post-sensing smoothing are jointly represented by a fixed exponential raw kernel
\begin{equation}
    g_{\mathrm{RX}}(t)
    =
    \frac{1}{\tau_{\mathrm{RX}}}
    \exp\left(-\frac{t}{\tau_{\mathrm{RX}}}\right),
    \qquad t\ge0,
    \label{eq:e2e_rx_kernel}
\end{equation}
and \(g_{\mathrm{RX}}(t)=0\) for \(t<0\). The normalized smoothing kernel used in \eqref{eq:e2e_observation} is \(r_{\mathrm{RX}}(t)=\mathcal N_t[g_{\mathrm{RX}}](t)\), with \(\tau_{\mathrm{RX}}=\qty{0.25}{\second}\).
The parameter \(\tau_{\mathrm{RX}}\) is an effective smoothing term and not an independently measured AS7341 time constant. We keep it fixed at \(\qty{0.25}{\second}\) as a compact modeling choice within the fitting procedure, so that \ac{RX} smoothing does not become another case-wise free parameter.

For a fixed source and kernel configuration, let \(m_j(\tau;s_j)\) denote the unscaled, time-shifted model waveform, and let \(\mathcal W_{\mathrm{early}}=\{\tau:\tau\le\qty{0.75}{\second}\}\). The nonnegative amplitude scale is fitted on the early part of the pulse by
\begin{equation}
    A_j(s_j)
    =
    \max\left\{
        0,\,
        \frac{
            \sum_{\tau\in\mathcal W_{\mathrm{early}}}
            y_j(\tau)m_j(\tau;s_j)
        }{
            \sum_{\tau\in\mathcal W_{\mathrm{early}}}
            m_j^2(\tau;s_j)
        }
    \right\}.
    \label{eq:e2e_scale_fit}
\end{equation}
The fitting objective intentionally emphasizes the dominant early pulse while retaining a tail penalty, because late tails are more sensitive to setup nonidealities and residual flow variations.

Model quality is evaluated with an objective that combines \ac{NRMSE} and \ac{CDFRMSE} terms while retaining a tail penalty,
\begin{equation}
    \begin{split}
    J_j
    &=
    0.80
    \left(
        \operatorname{NRMSE}_{\mathrm{early},j}
        +
        0.25\,\operatorname{CDFRMSE}_{\mathrm{early},j}
    \right)
    \\
    &\quad
    +
    0.20\,\operatorname{NRMSE}_{\mathrm{tail},j},
    \end{split}
    \label{eq:e2e_objective}
\end{equation}
Here, the fixed early window is \(\tau\le\qty{0.75}{\second}\), and the fixed tail window is \(\tau\ge\qty{1.0}{\second}\). The \ac{NRMSE} values are normalized by the measured peak of the corresponding case. The \ac{CDFRMSE} term is the root-mean-square error between the cumulative sums of the nonnegative measured and modeled early-window curves, again normalized consistently within each case; the full metric definitions are reported in the supplementary material.

Source-shape parameters are regularized toward color-specific centers so that within-color release profiles remain similar. The exact transformed parameter vector, regularization term, and color centers are reported in the supplementary material.

The \ac{AD}-style fitting yields the global grid-selected value \(B_{\mathrm{eff}}=1.1\times10^{-5}\,\si{\meter\squared\per\second}\), while for the inner-disk benchmark we obtain the globally profiled value \(\alpha=r_i/r_c=0.30\). These global selections are applied across all colors and distances, showing that one compact propagation setting remains useful beyond a single isolated \ac{CIR}. The fitted \(\alpha=0.30\) is plausible for a localized side-needle release occupying only part of the tube cross-section and broadly aligns with the testbed geometry and visual intuition, but it is not a measured plume radius. Similarly, \(B_{\mathrm{eff}}\) remains a fitted effective broadening coefficient, not a directly measured diffusion parameter. Table~\ref{tab:e2e_parameter_policy} summarizes the most important fixed, fitted, and nuisance parameters, while a more detailed supplementary table lists the full parameter grids and selected color-regularization centers.

\begin{figure*}[t]
    \centering
    \includegraphics[width=0.8\linewidth]{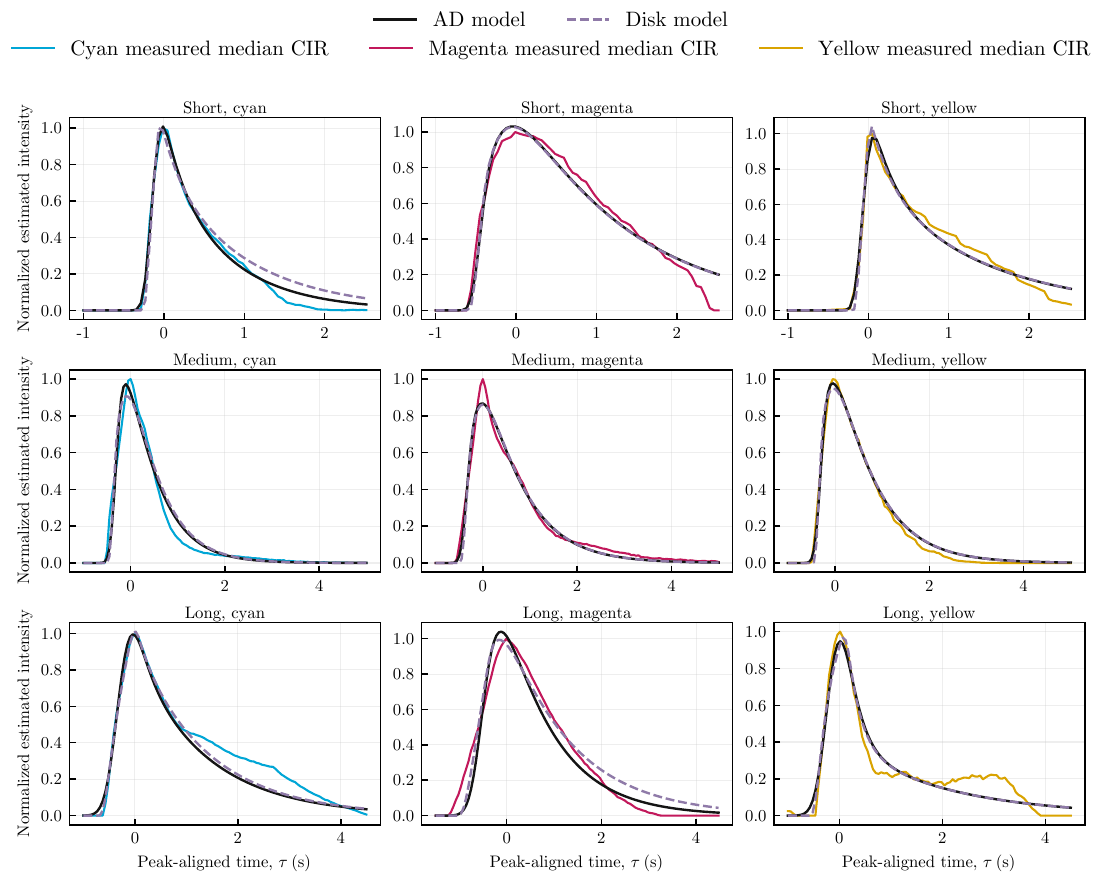}
    \caption{Peak-normalized measured median \acp{CIR} and effective-model overlays at the selected distances. Rows correspond to the short, medium, and long modeling distances, and columns correspond to cyan, magenta, and yellow. Each measured curve is the pointwise median across the retained, peak-aligned pulses. The measured median and both model predictions are normalized by the peak of that measured median. The \ac{AD}-style and inner-disk candidates use the common finite-release, receiver-smoothing, amplitude-scaling, and timing-alignment framework.}
    \label{fig:ad_vs_disk_grid}
\end{figure*}

\begin{table}[t]
    \centering
    \caption{Compact parameter policy for the effective end-to-end model used in the main text.}
    \label{tab:e2e_parameter_policy}
    \setlength{\tabcolsep}{3.5pt}
    \resizebox{\linewidth}{!}{%
    \begin{tabular}{p{0.15\textwidth} p{0.275\textwidth} p{0.25\textwidth}}
        \toprule
        \textbf{Symbol} & \textbf{Role / status} & \textbf{Value} \\
        \midrule
        \(\ell\) & distance metadata (fixed) & \(\qty{9.5}{\centi\meter}\), \(\qty{23.5}{\centi\meter}\), \(\qty{100}{\centi\meter}\) \\
        \(Q_0,r_c\) & hardware metadata (fixed) & \(Q_0=\qty{10}{\milli\liter\per\minute}\), \(r_c=\qty{0.8}{\milli\meter}\) \\
        \(v_{\mathrm{avg}}\) & mean speed (derived) & \(Q_0/(\pi r_c^2)\) \\
        \(B_{\mathrm{eff}}\) & \ac{AD} broadening coefficient (global fit) & \(1.1\times10^{-5}\,\si{\meter\squared\per\second}\) \\
        \(\alpha=r_i/r_c\) & disk support (global fit) & \(0.30\) \\
        \(\ell_{\mathrm{RX}}\) & \ac{RX} window (fixed modeling choice) & \(\qty{5}{\milli\meter}\) \\
        \(\tau_{\mathrm{RX}}\) & \ac{RX} smoothing (fixed modeling choice) & \(\qty{0.25}{\second}\) \\
        \(\lambda_j,\tau_{\mathrm{build},j},\tau_{\mathrm{decay},j}\) & source shape (case-wise fitted) & grids in supplementary material \\
        \(A_j\) & response amplitude (case-wise analytic fit) & nonnegative least-squares fit; Eq. \eqref{eq:e2e_scale_fit} \\
        \(s_j\) & timing alignment (case-wise grid fit) & grid in supplementary material \\
        \(\rho\) & regularization strength (global fit) & \ac{AD} \(0.05\), disk \(0.03\) \\
        \bottomrule
    \end{tabular}%
    }
\end{table}

\subsection{Modeling Results and Diagnostic Limitations}
\label{subsec:modeling_results_limits}

Fig.~\ref{fig:ad_vs_disk_grid} compares the measured \acp{CIR} with the \ac{AD}-style fit and the inner-disk benchmark at the selected distances. All curves are peak-normalized per panel, so the figure should be interpreted primarily as a shape comparison rather than an amplitude comparison.

The main qualitative result is that both compact end-to-end models capture much of the rise, peak, and dominant decay behavior across colors and distances. The overlays are strongest for the short- and medium-distance cases, where the dominant timing and pulse width align well after finite release, propagation/residence, \ac{RX} smoothing, amplitude scaling, and timing alignment are fitted. This supports the use of finite-release source modeling combined with flow-scale transport, but it does not imply that all residual tail behavior is explained.

The aggregate metrics reported in the supplementary material support this visual comparison. Reported as the mean \(\pm\) sample standard deviation across all nine color-distance cases, the overall objective is \(0.0523\pm0.0223\) for the \ac{AD}-style model and \(0.0559\pm0.0175\) for the inner-disk model; the \ac{AD}-style value is lower at the short and medium distances, whereas the inner-disk value is lower at the long distance.

The color-dependent mismatches are also informative. In particular, magenta shows a delayed leakage/tail effect, especially in the short-distance \ac{CIR}, which may be related to practical needle interaction in the \ac{TX} arrangement. Across colors, late tails are generally less accurate than the rise, peak, and dominant decay. This follows partly from the fitting objective, which intentionally prioritizes the dominant early pulse, and partly from the fact that tails are more affected by pump variability, residual flow fluctuations, unmodeled dispersion/noise, and spectral-estimation effects. The lower long-distance agreement may reflect the accumulation of model mismatch and experimental nonidealities over the longer propagation distance.

The \ac{AD}-style model provides a smooth effective broadening description with one global \(B_{\mathrm{eff}}\), while the inner-disk model provides a laminar-flow benchmark that tests whether a finite local radial support can explain similar residence-time behavior without introducing an effective diffusion-like parameter. Overall, the modeling section supports three safe claims: the dominant \ac{CIR} timing is flow-scale, finite release is necessary for an interpretable end-to-end response, and compact \ac{AD}-style and inner-disk kernels provide useful complementary descriptions. It does not support claims of literal molecular diffusion identification or exact injection recovery. The supplement reports the full fitting definitions and parameter grids together with distance-resolved fitted source profiles, measured-only \acp{CIR}, and selected residual and late-tail diagnostics.

\section{Detector Benchmark on Real \texorpdfstring{\acs{MUMO}}{MUMO} Data}
\label{sec:detector_benchmark}

This section compares practical detector families using measured payload traces and acquired \ac{CIR} measurements over several thousand transmitted bits. The benchmark covers two distances, trace-only and \ac{CIR}-assisted methods, and low- and high-complexity detection approaches.
Detector performance is evaluated by \ac{BER} under optimized detector parameters for each operating point, so the results compare the best fit of each detector class on the available measurement data.

\subsection{Operating Points and Benchmark Protocol}
\label{subsec:detector_protocol}

We evaluate the measured continuous \ac{MUMO}-\ac{OOK} payload traces obtained from the spectral \ac{RX} processing in Sec.~\ref{subsec:testbed_rx_spectral_estimation}. Each color $c$ in \(\mathcal C=\{\mathrm C,\mathrm M,\mathrm Y\}\) is treated as one binary \ac{OOK} subchannel. Unless the text explicitly discusses pooled benchmark accounting, the notation fixes one operating point and one color. The estimated trace is written as \(y(t)\), the transmitted bit as \(b_k\in\{0,1\}\), and the decision of detector \(d\) as \(\hat b_k^{(d)}\in\{0,1\}\). The fitting itself remains operating-point-specific but the compact notation avoids carrying these bookkeeping indices through every detector equation.

The evaluated operating points use the channel lengths \(\ell=\qty{7.4}{\centi\meter}\) and \(\ell=\qty{23.5}{\centi\meter}\), denoted as \textit{short-} and \textit{medium-}channel scenarios. The symbol durations \(T_{\mathrm{sym}}\in\{\qty{0.5}{\second},\qty{0.667}{\second},\qty{1}{\second},\qty{2}{\second}\}\), and injection durations \(T_{\mathrm{inj}}\in\{\qty{0.05}{\second},\qty{0.1}{\second},\qty{0.2}{\second}\}\), are applied depending on the measurement campaign, as listed in Table~\ref{tab:detector_operating_points}. 

We separate the measured operating points into \textit{Main Settings} and \textit{Stress Settings}. \textit{Main Settings} describe the intended operating envelope after manually selecting testbed hardware parameters such as \(T_{\mathrm{inj}}\), while \textit{Stress Settings} deliberately retain deviations from those choices to show unfavorable but informative injection-duration and high-rate regimes.
The medium-distance payloads use aggregate-balanced 100-symbol sequences, while the short-distance payloads use per-color balanced 300-symbol sequences. Overall, the \textit{Main Settings} contain a balanced set of \(4500\) evaluated payload bits, the \textit{Stress Settings} contain \(2100\) evaluated payload bits, and the combined benchmark contains \(6600\) evaluated payload bits. These counts exclude preamble symbols.

When all three color subchannels are active, one \ac{MUMO} symbol carries three binary bits. The nominal aggregate bit rate is therefore
\begin{equation}
    R_{\mathrm{agg}}
    =
    \frac{3}{T_{\mathrm{sym}}}.
    \label{eq:detector_ragg}
\end{equation}

For each operating point, color, and detector variant, timing, threshold, template, and detector-specific parameters are optimized on the complete available payload sequence. The resulting \ac{BER} values are measurement-optimized benchmark values that quantify the best fit of each detector class to the measured traces. These values do not imply out-of-sample generalization.

For scalar-threshold detectors at a fixed operating point and color, the symbol starts for \(k=0,\ldots,M-1\) are modeled as
\begin{equation}
    t_k
    =
    t_0
    +
    k\left(T_{\mathrm{sym}}+\Delta\right),
    \label{eq:detector_symbol_starts}
\end{equation}
where \(t_0\) is the optimized detector start time and \(\Delta\) is an empirically fitted processing-delay correction for the microcontroller-driven symbol schedule with an average value of \(\Delta_\mathrm{avg}=\qty{23.75}{\milli\second}\). A nominal detector window is defined as
\begin{equation}
    \mathcal W_k^{(d)}
    =
    \left[
        t_k,\,
        t_k+T_{\mathrm{sym}}
    \right).
    \label{eq:detector_symbol_window}
\end{equation}
For symbol-by-symbol detectors, the samples in this window form a scalar statistic \(z_k^{(d)}\). The implemented decision and peak-search windows both span this complete symbol interval, and the energy statistics use the same samples without an independent energy subwindow. The supplement specifies the corresponding stored-sample and local-reference conventions.

Scalar-threshold detectors use the common hard-decision rule
\begin{equation}
    \hat b_k^{(d)}
    =
    \begin{cases}
        1, & z_k^{(d)}\ge\tau^{(d)},\\
        0, & \text{otherwise},
    \end{cases}
    \label{eq:detector_threshold_rule}
\end{equation}
where \(\tau^{(d)}\) is the optimized threshold. Let \(\mathcal P\) denote the payload-symbol index set of the fixed operating point, excluding preamble symbols. The optimized parameter vector is selected by minimizing the number of bit errors over this set,
\begin{equation}
    \theta_d^{\star}
    =
    \arg\min_{\theta\in\Theta_d}
    \left|
        \{k\in\mathcal P:\hat b_k^{(d)}(\theta)\ne b_k\}
    \right|,
    \label{eq:detector_oracle_objective}
\end{equation}
where \(\theta\) collects the active timing, threshold, window, reference, feedback, template, or memory parameters of detector \(d\), and \(\Theta_d\) is the corresponding candidate set. This optimization is repeated for each detector and operating point in the full benchmark.

For pooled reporting, let \(\mathcal P_o\) denote the payload-symbol index set of operating point \(o\) for one color, and let \(E_{c,o}^{(d)}\) be the number of bit errors made by detector \(d\) for color \(c\) at operating point \(o\). The pooled \ac{BER} over a set of operating points \(\mathcal O\) is then
\begin{equation}
    \mathrm{BER}^{(d)}
    =
    \frac{
        \sum_{o\in\mathcal O}
        \sum_{c\in\mathcal C}
        E_{c,o}^{(d)}
    }{
        \sum_{o\in\mathcal O}
        \sum_{c\in\mathcal C}
        |\mathcal P_o|
    }.
    \label{eq:detector_ber}
\end{equation}
We also report an empirical \ac{MI}-based decision-information-rate proxy
\begin{equation}
    R_{\mathrm{MI},o}^{(d)}
    =
    \frac{
        \hat I_{o}^{(\mathrm C,d)}
        +
        \hat I_{o}^{(\mathrm M,d)}
        +
        \hat I_{o}^{(\mathrm Y,d)}
    }{
        T_{\mathrm{sym},o}
    },
    \label{eq:detector_rmi}
\end{equation}
where \(\hat I_{o}^{(c,d)}\) is estimated from the empirical confusion matrix of color \(c\) after applying detector \(d\). For transmitted bit \(X\) and detector decision \(\hat X\), this estimate is
\begin{equation}
    \hat I_{o}^{(c,d)}
    =
    \sum_{i\in\{0,1\}}
    \sum_{j\in\{0,1\}: \hat p_{ij}>0}
    \hat p_{ij}
    \log_2
    \left(
        \frac{\hat p_{ij}}{\hat p_{i\cdot}\hat p_{\cdot j}}
    \right),
    \label{eq:detector_empirical_mi}
\end{equation}
where \(\hat p_{ij}\) is the empirical probability of \(X=i\) and \(\hat X=j\), zero joint-probability terms are omitted, and the balanced payload construction avoids zero input marginals in the evaluated cases. For example, in the short-channel, \(T_{\mathrm{sym}}=\qty{0.5}{\second}\), \(T_{\mathrm{inj}}=\qty{0.2}{\second}\) \textit{Main Setting}, \ac{MEDD} makes no errors over the \(900\) evaluated bits. The resulting empirical information is one bit per color decision and \(R_{\mathrm{MI}}=\qty{6}{\bit\per\second}\). This quantity describes detector-level decision information after measurement data fitting and should not be interpreted as a channel capacity. Table~\ref{tab:detector_operating_points} lists the evaluated operating points and gives the best detector per condition for orientation before the family-level comparison.

\begin{table}[t]
    \centering
    \caption{Detector benchmark operating points grouped into
    \textit{Main Settings} and \textit{Stress Settings}. Short and Medium
    denote the \(\qty{7.4}{\centi\meter}\) and \(\qty{23.5}{\centi\meter}\)
    payload channels, respectively. A dagger denotes a tie within the
    corresponding trace-only or \acs{CIR}-assisted class.}
    \label{tab:detector_operating_points}
    \setlength{\tabcolsep}{4pt}
    \resizebox{\linewidth}{!}{%
    \begin{tabular}{
        p{0.06\textwidth}
        p{0.06\textwidth}
        p{0.06\textwidth}
        p{0.08\textwidth}
        p{0.18\textwidth}
        p{0.14\textwidth}}
        \toprule
        \textbf{Length} &
        \(\boldsymbol{T_{\mathrm{sym}}}\) &
        \(\boldsymbol{T_{\mathrm{inj}}}\) &
        \(\boldsymbol{R_{\mathrm{agg}}}\) &
        \textbf{Best detector} &
        \textbf{\acs{BER}} \\
        & & & &
        \multicolumn{1}{l}{\scriptsize Trace-only \(\mid\) \acs{CIR}-assisted} &
        \multicolumn{1}{l}{\scriptsize Trace-only \(\mid\) \acs{CIR}-assisted} \\
        \midrule
        \multicolumn{6}{l}{\textbf{Main Settings}} \\
        \midrule

        Medium &
        \(\qty{0.5}{\second}\) &
        \(\qty{0.05}{\second}\) &
        \(\qty{6.0}{\bit\per\second}\) &
        \ac{MEDD} \(\mid\) \ac{MMSE} &
        \(11.33\% \mid 10.67\%\) \\

        Medium &
        \(\qty{1.0}{\second}\) &
        \(\qty{0.05}{\second}\) &
        \(\qty{3.0}{\bit\per\second}\) &
        \ac{MEDD} \(\mid\) \ac{MMSE} &
        \(0.00\% \mid 0.00\%\) \\

        Medium &
        \(\qty{2.0}{\second}\) &
        \(\qty{0.10}{\second}\) &
        \(\qty{1.5}{\bit\per\second}\) &
        \ac{MEDD}\textsuperscript{\(\dagger\)}
        \(\mid\)
        \ac{MMSE}\textsuperscript{\(\dagger\)} &
        \(0.00\% \mid 0.00\%\) \\

        Short &
        \(\qty{0.5}{\second}\) &
        \(\qty{0.20}{\second}\) &
        \(\qty{6.0}{\bit\per\second}\) &
        \ac{MEDD} \(\mid\) \ac{MMSE} &
        \(0.00\% \mid 0.00\%\) \\

        Short &
        \(\qty{0.667}{\second}\) &
        \(\qty{0.10}{\second}\) &
        \(\qty{4.50}{\bit\per\second}\) &
        \ac{MEDD} \(\mid\) \ac{MMSE} &
        \(0.00\% \mid 0.00\%\) \\

        Short &
        \(\qty{1.0}{\second}\) &
        \(\qty{0.20}{\second}\) &
        \(\qty{3.0}{\bit\per\second}\) &
        \ac{MEDD}\textsuperscript{\(\dagger\)}
        \(\mid\)
        \ac{MMSE}\textsuperscript{\(\dagger\)} &
        \(0.00\% \mid 0.00\%\) \\

        Short &
        \(\qty{2.0}{\second}\) &
        \(\qty{0.20}{\second}\) &
        \(\qty{1.5}{\bit\per\second}\) &
        \ac{MEDD}\textsuperscript{\(\dagger\)}
        \(\mid\)
        \ac{MMSE}\textsuperscript{\(\dagger\)} &
        \(0.00\% \mid 0.00\%\) \\

        \midrule
        \multicolumn{6}{l}{\textbf{Stress Settings}} \\
        \midrule

        Short &
        \(\qty{0.5}{\second}\) &
        \(\qty{0.10}{\second}\) &
        \(\qty{6.0}{\bit\per\second}\) &
        \ac{MEDD} \(\mid\) \ac{MMSE} &
        \(21.22\% \mid 20.72\%\) \\

        Medium &
        \(\qty{1.0}{\second}\) &
        \(\qty{0.10}{\second}\) &
        \(\qty{3.0}{\bit\per\second}\) &
        \ac{MEDD} \(\mid\) \ac{MMSE} &
        \(3.00\% \mid 4.00\%\) \\

        \bottomrule
    \end{tabular}%
    }
\end{table}

\subsection{Decision Statistic on a Measured Trace}
\label{subsec:detector_decision_trace}

\begin{figure*}[t]
    \centering
    \includegraphics[width=0.7\linewidth]{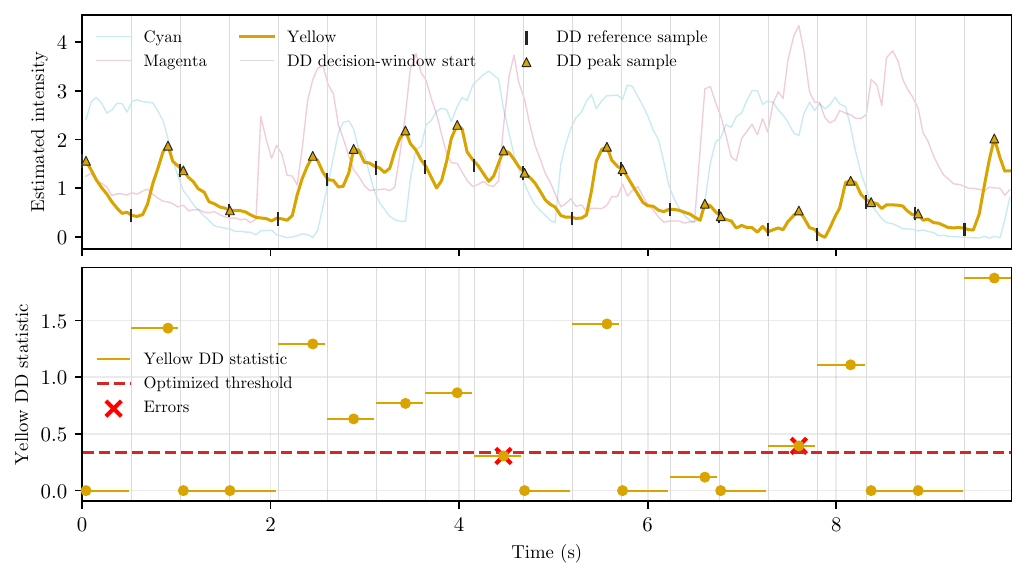}
    \caption{Yellow-channel \acs{DD} decision mechanism for the short-channel setting with symbol period \(T_{\mathrm{sym}}=\qty{0.5}{\second}\) and injection duration \(T_{\mathrm{inj}}=\qty{0.2}{\second}\). The upper panel shows the measured color traces, decision-window starts, local reference samples, and selected yellow-channel peaks. The lower panel shows the resulting peak-minus-reference statistic and optimized threshold; crosses mark decision errors. The other color traces are faded to retain the simultaneous \acs{MUMO} context. The \acs{DD} scheme is chosen because its metric calculation can be easily visualized.}
    \label{fig:dd_decisions}
\end{figure*}

Before introducing the full detector taxonomy, Fig.~\ref{fig:dd_decisions} shows how a scalar detector statistic is applied to a real measured \ac{MUMO} payload trace. The example uses the yellow channel of the short-channel, \(T_{\mathrm{sym}}=\qty{0.5}{\second}\), \(T_{\mathrm{inj}}=\qty{0.2}{\second}\) \textit{Main Setting}. The upper panel marks the local reference and peak samples selected from each decision window, while the lower panel shows the resulting \ac{DD} statistic, optimized threshold, and decision errors. This example provides a concrete reference for the subsequent detector definitions. Each detector converts a symbol-aligned portion of the received trace into a scalar score or sequence score, which is then compared with a fitted decision rule under the benchmark protocol.

\subsection{Detector Families}
\label{subsec:detector_families}


\definecolor{ModelFreeGray}{gray}{0.94}
\definecolor{CIRGray}{gray}{0.90}

\begin{table*}[t]
\caption{Detector variants used in the benchmark.}
\label{tab:detector_taxonomy}
\centering
\resizebox{\textwidth}{!}{%
\begin{tabular}{p{0.23\textwidth} p{0.12\textwidth} p{0.47\textwidth} p{0.11\textwidth}}
\toprule
\textbf{Plot label} &
\textbf{Family} &
\textbf{Core idea} &
\textbf{Memory} \\
\midrule

\rowcolor{ModelFreeGray}
\multicolumn{4}{@{}l}{\textit{Trace-only}: denotes detectors that use only measured received-signal statistics.} \\
\midrule

\acs{DFDD}
& Decision-feedback
& Feedback-corrected decision statistic
& Decisions \\

\acs{DD}
& Difference
& Peak-minus-start difference
& -- \\

\acs{EDD}
& Difference
& Mean-normalized energy difference with sequence-start zero initialization
& Energy \\

\acs{MEDD}
& Difference
& Adaptive energy-difference statistic
& Previous energy \\

Energy mean
& Energy
& Window-mean intensity
& -- \\

Energy mean, baseline-sub.
& Energy
& Baseline-subtracted window-mean intensity
& -- \\

Energy mean, clipped baseline-sub.
& Energy
& Clipped baseline-subtracted window-mean intensity
& -- \\

\midrule
\rowcolor{CIRGray}
\multicolumn{4}{@{}l}{\textit{\acs{CIR}-assisted}: denotes detectors that use estimated channel impulse responses.} \\
\midrule

\acs{MLSD} Viterbi
& \acs{MLSD}
& Viterbi sequence search using channel memory
& Sequence \\

\acs{MLSD} simplified
& \acs{MLSD}
& Greedy cancellation using channel memory
& Sequence \\

\acs{MMSE}
& \acs{MMSE}
& Full-sequence regularized convolution solve
& Equalizer \\

\acs{MF}
& \acs{MF}
& \ac{CIR}-correlation statistic
& Symbol \\

        \bottomrule
\end{tabular}%
}
\end{table*}

Table~\ref{tab:detector_taxonomy} summarizes the evaluated detector variants and separates trace-only detectors from \ac{CIR}-assisted ones. Energy-type detectors provide low-complexity threshold baselines based on integrated or averaged window intensity~\cite{llatserDetectionTechniquesDiffusionbased2013,kilincReceiverDesignMolecular2013,qianMolecularCommunicationsModelBased2019a}, with baseline-subtracted and clipped variants reducing sensitivity to slow drift. Difference-detection variants compare a current symbol statistic to a local or previous-symbol reference, following the practical direction of adaptive, memory-assisted, and noncoherent designs for scenarios where stable channel knowledge is difficult to maintain~\cite{damrathLowComplexityAdaptiveThreshold2016,alshammriLowcomplexityMemoryassistedAdaptivethreshold2017,noelAsynchronousPeakDetection2017,wietfeldEvaluationMultiMoleculeMolecular2024b}.

\ac{CIR}-assisted detectors provide model-informed comparison points when channel templates are available~\cite{kilincReceiverDesignMolecular2013,mengReceiverDesignDiffusionBased2014,qianMolecularCommunicationsModelBased2019a}. Matched-filter variants correlate observations with static \ac{CIR} templates~\cite{jamaliDesignMatchedFilters2017}, while \ac{MMSE} and \ac{MLSD} variants use finite-memory \ac{CIR} representations for equalization or sequence scoring.

The optimized parameter classes include timing, thresholds, feedback memory, \ac{MEDD} scaling and offset, equalizer regularization, and sequence memory. The detector and peak windows are fixed to the full symbol interval. Exact implementation details, full parameter grids, and variant-specific settings are deferred to the supplement.

\subsection{Trace-Only Difference Statistics}
\label{subsec:trace_only_difference_statistics}

For compact notation in this subsection, the operating point and color are fixed and the indices \(o\) and \(c\) are omitted. The conference precursor used a simple difference-detection statistic that compares the peak in a symbol window to a local reference~\cite{wietfeldEvaluationMultiMoleculeMolecular2024b}. This local-reference idea is related to adaptive and asynchronous peak-detection \ac{RX}s that avoid relying on a perfectly stable absolute concentration baseline~\cite{damrathLowComplexityAdaptiveThreshold2016,noelAsynchronousPeakDetection2017}. The peak-search interval is \(\mathcal T_k^{\mathrm{peak}}=[t_k,t_k+T_{\mathrm{sym}})\), and the local reference \(r_k\) is the first received sample at or after the fitted symbol start \(t_k\). The implemented \ac{DD} statistic is the absolute peak-minus-reference difference
\begin{equation}
    z_k^{(\mathrm{DD})}
    =
    \left|
        \max_{t\in\mathcal T_k^{\mathrm{peak}}} y(t)
        -
        r_k
    \right|.
    \label{eq:dd_detector}
\end{equation}
This makes \ac{DD} sensitive to local peaks, but it can also be affected by peak timing errors and residual tails from previous symbols.

\begin{figure}[t]
    \centering
    \includegraphics[width=0.7\linewidth]{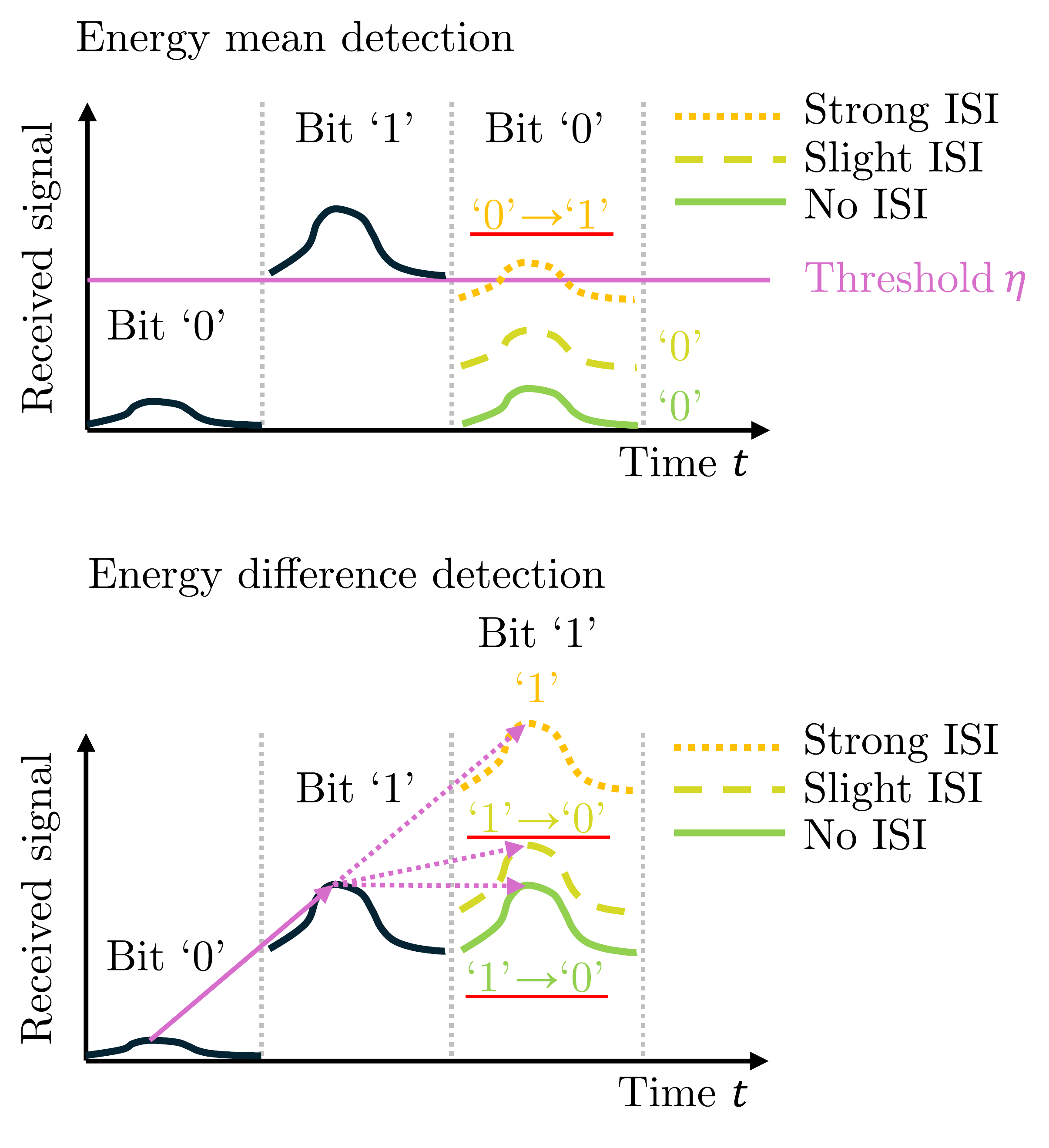}
    \caption{Schematic failure modes for energy-mean and energy-difference decisions. The top panel illustrates how \ac{ISI} can cause an energy-threshold false positive for a zero symbol. The bottom panel illustrates how subtraction of a large previous-symbol reference can cause an energy-difference false negative for a one symbol. Solid, dashed, and dotted curves consistently denote no, slight, and strong \acs{ISI}, respectively, while an underline denotes a wrong resulting bit decision.}
    \label{fig:energy_edd_failure_modes}
\end{figure}

\ac{EDD} instead compares consecutive symbol energies. It can be interpreted as a trace-only energy-threshold statistic with a previous-symbol reference, following the broader class of low-complexity and memory-assisted \ac{MC} detectors~\cite{llatserDetectionTechniquesDiffusionbased2013,alshammriLowcomplexityMemoryassistedAdaptivethreshold2017,qianMolecularCommunicationsModelBased2019a}. We use the mean-normalized window energy
\begin{equation}
    e_k
    =
    \frac{1}{|\mathcal W_k^{(d)}|}
    \sum_{m\in\mathcal W_k^{(d)}}y[m],
    \label{eq:edd_energy}
\end{equation}
which prevents the statistic scale from changing solely because candidate windows contain different numbers of samples. The \ac{EDD} statistic is
\begin{equation}
    z_k^{(\mathrm{EDD})}
    =
    e_k
    -
    e_{k-1}.
    \label{eq:edd_detector}
\end{equation}
The previous energy is initialized to zero only at the first symbol of the complete preamble-plus-payload sequence. Consequently, the first evaluated payload symbol uses the energy of the final preamble symbol. A large previous symbol, a weak current symbol, or slow baseline drift can nevertheless distort the raw difference. Fig.~\ref{fig:energy_edd_failure_modes} illustrates the corresponding failure modes for energy-mean and energy-difference decisions under residual tails and \ac{ISI}: a residual tail can lift a zero above an energy threshold, while subtraction from a large previous energy can suppress a following one. The schematic motivates the adaptive-reference modification introduced next, but it does not depict the \ac{MEDD} statistic directly.

We introduce \ac{MEDD} in this paper as a small adaptive-reference modification of \ac{EDD}. Instead of subtracting the raw previous energy directly, \ac{MEDD} scales the previous reference and adds an offset based on the standard deviation of the payload-symbol energies in the corresponding physical trace,
\begin{equation}
    \begin{aligned}
    z_k^{(\mathrm{MEDD})}
    &=
    e_k
    -A e_{k-1}
    +B_{\mathrm{off}},
    \\
    B_{\mathrm{off}}
    &=
    B_{\mathrm{scale}}
    \operatorname{std}_{n\in\mathcal P}(e_n).
    \end{aligned}
    \label{eq:medd_detector}
\end{equation}
The same full-sequence initialization is used for \ac{MEDD}: only the first preamble-plus-payload symbol has a zero previous-energy reference, while the first evaluated payload symbol uses the final preamble energy. \ac{MEDD} remains trace-only and keeps the same basic symbol-by-symbol structure as \ac{EDD}. The additional parameters \(A\) and \(B_{\mathrm{scale}}\) reduce the impact of residual tails, baseline drift, and alternating pulse amplitudes on the previous-symbol reference. Importantly, \ac{MEDD} does not use hard-decision feedback.

As a separate decision-feedback comparator, \ac{DFDD} starts from the \ac{DD} statistic and subtracts a short-memory correction based on previous hard decisions,
\begin{equation}
    \begin{aligned}
    z_k^{(\mathrm{DFDD})}
    &=
    z_k^{(\mathrm{DD})}
    -
    \gamma
    \sum_{i=1}^{\min(L,k)}
    \hat b_{k-i},
    \\
    \hat b_k
    &=
    \begin{cases}
        1, & z_k^{(\mathrm{DFDD})}\ge\eta,\\
        0, & \text{otherwise}.
    \end{cases}
    \end{aligned}
    \label{eq:dfdd_detector}
\end{equation}
Here, \(\hat b_{k-i}\) denotes the detector's own previous hard decisions, \(\gamma\) is the feedback coefficient, \(L\) is the decision-feedback memory, and \(\eta\) is the decision threshold; these quantities are selected by the benchmark optimization protocol.

\subsection{\texorpdfstring{\acs{CIR}}{CIR}-Assisted Detector Statistics}
\label{subsec:cir_assisted_detector_statistics}

The \ac{CIR}-assisted detectors use symbol-spaced or windowed templates derived from the available measured \acp{CIR}. They compare the trace-only statistics against \ac{RX} classes that use explicit channel-response information, while preserving the measurement-optimized fitting policy of the benchmark. As in the previous subsection, the operating point and color are fixed in the notation below.

The matched-filter detector forms a forward-looking correlation between the symbol-spaced observation sequence \(u_k\) and a selected \ac{CIR} template \(h_q\). The reported \ac{MF} uses a fixed color-matched \ac{CIR} template policy: each observed color is paired with a separate measured template for that color, and the assignment is reused across the relevant payload conditions. With squared-template normalization, the implemented score can be written as
\begin{equation}
    z_k^{(\mathrm{MF})}
    =
    \frac{
        \sum_{q=0}^{L_{\mathrm{MF}}-1}u_{k+q}h_q
    }{
        \sum_{q=0}^{L_{\mathrm{MF}}-1}h_q^2
    },
    \label{eq:mf_detector}
\end{equation}
followed by the scalar threshold rule in \eqref{eq:detector_threshold_rule}~\cite{jamaliDesignMatchedFilters2017}.

The \ac{MMSE} detector uses a finite-memory convolution representation of the symbol-spaced \ac{CIR}. Let \(\mathbf y\) denote the observed scalar sequence after the selected observation mapping, and let \(\mathbf H\) be the lower-triangular convolution matrix formed from the selected \ac{CIR} taps. The implemented benchmark solves the regularized least-squares problem
\begin{equation}
    \hat{\mathbf x}
    =
    \left(
        \mathbf H^{\mathsf T}\mathbf H
        +
        \lambda_{\mathrm{MMSE}}\mathbf I
    \right)^{-1}
    \mathbf H^{\mathsf T}\mathbf y,
    \label{eq:mmse_detector}
\end{equation}
and thresholds the resulting entries \(\hat x_k\).

The \ac{MLSD} variants use the finite-memory sequence model
\begin{equation}
    y_k
    \approx
    \sum_{q=0}^{L-1}
    h_q b_{k-q},
    \label{eq:mlsd_sequence_model}
\end{equation}
where \(h_q\) are symbol-spaced \ac{CIR} taps. The initialization of both implemented variants is specified in the supplement. Using an effective variance scale \(\sigma^2\), the corresponding sequence decision can be written as
\begin{equation}
    \hat{\mathbf b}
    =
    \arg\min_{\mathbf b\in\{0,1\}^{M}}
    \sum_{k}
    \frac{
        \left(
            y_k
            -
            \sum_{q=0}^{L-1}
            h_q b_{k-q}
        \right)^2
    }{\sigma^2},
    \label{eq:mlsd_metric}
\end{equation}
where \(M\) is the number of payload symbols in the considered sequence. \(\sigma^2\) is an effective metric scale computed from the variance of the measurement samples \(y_k\) for the considered operating point and color. To make the implementation practical, the memory length is restricted to \(L\) and retained paths to \(P\). The simplified \ac{MLSD} variant greedily subtracts predicted previous-symbol \ac{ISI}, while the Viterbi variant uses a pruned binary finite-state recursion inspired by classical sequence detection~\cite{forneyMaximumLikelihoodSequence1973,viterbiErrorBoundsAsymptotically1967}.

Table~\ref{tab:detector_complexity} summarizes the order-level operating complexity after the detector parameters are specified. The trace-only energy, \ac{EDD}, \ac{MEDD}, and \ac{DD} variants require one pass over the samples in each symbol window and therefore scale as \(O(MN)\). \ac{DFDD} adds a short hard-decision feedback memory, while \ac{MF} additionally evaluates a symbol-spaced correlation of length \(L_{\mathrm{MF}}\).
The model-assisted sequence and equalization benchmarks are more expensive and depend more strongly on the assumed memory length. The \ac{MMSE} benchmark uses a dense regularized full-sequence solve, whereas \ac{MLSD} ranges from a greedy \(O(ML)\) approximation to the implemented \(O(MPL)\) pruned Viterbi search. The exhaustive binary-state reference is discussed in the supplement.

\begin{table}[t]
    \centering
    \caption{Detector complexity for one fixed parameter setting, where \(M\) is the number of payload symbols per color, \(N\) the samples per detector window, \(L\) the detector memory, \(L_{\mathrm{MF}}\) the matched-filter length, and \(P\) the retained Viterbi paths. The \acs{MMSE} entry describes the dense full-sequence implementation; derivations are reported in the supplement.}
    \label{tab:detector_complexity}
    \scriptsize
    \setlength{\tabcolsep}{3pt}
    \begin{tabular}{p{0.35\linewidth} p{0.50\linewidth}}
        \toprule
        \textbf{Detector class} & \textbf{Operating complexity} \\
        \midrule
        Energy, \ac{EDD}, \ac{MEDD}, \ac{DD} & \(O(MN)\) \\
        \ac{DFDD} & \(O(M(N+L))\) \\
        \ac{MF} & \(O(M(N+L_{\mathrm{MF}}))\) \\
        \ac{MMSE} benchmark & dense \(O(M^3)\) solve and \(O(M^2)\) memory \\
        \ac{MLSD} simplified & \(O(ML)\) \\
        \ac{MLSD} Viterbi, \(P=8\) & \(O(MPL)\) \\
        \bottomrule
    \end{tabular}
\end{table}

\subsection{Detector-Family Comparison}
\label{subsec:detector_family_comparison}

\begin{figure}[t]
    \centering
    \includegraphics[width=0.9\linewidth]{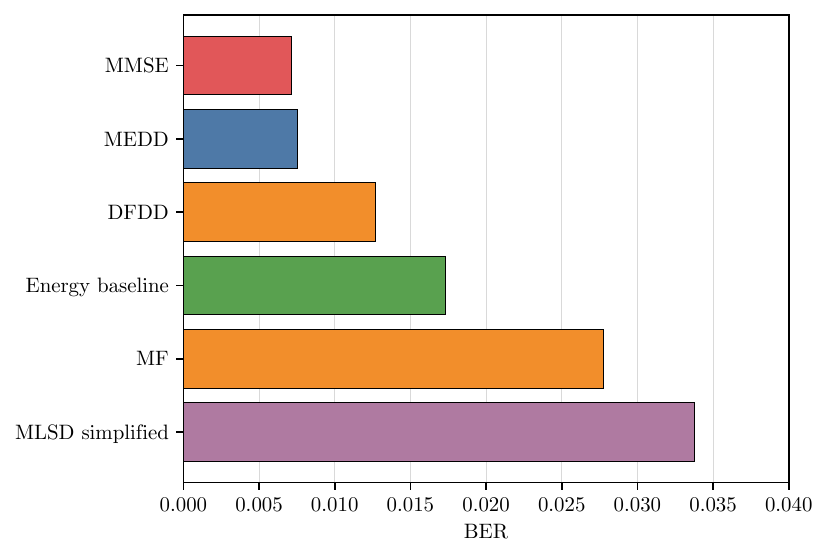}
    \caption{Detector family-level \acs{BER} comparison using one fixed detector variant selected per family over the \textit{Main Settings}. The benchmark is optimized using measurements of the available payload traces.}
    \label{fig:detector_family_comparison}
\end{figure}

\begin{figure}[t]
    \centering
    \includegraphics[width=0.8\linewidth]{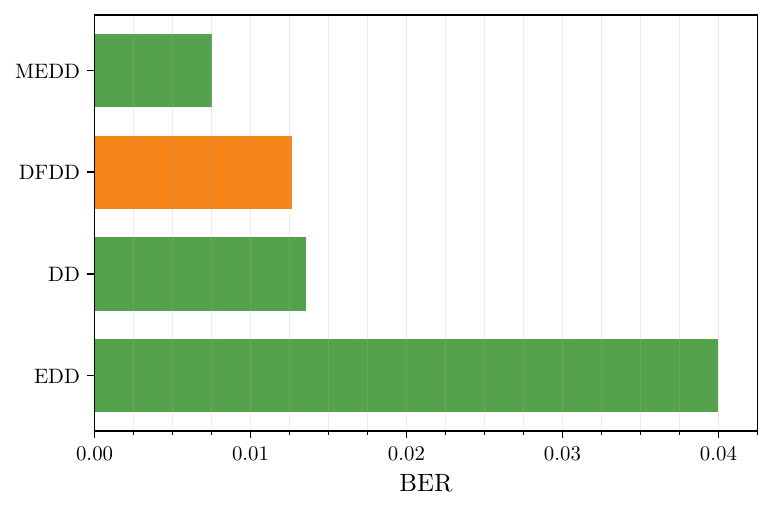}
    \caption{Trace-only difference-detection variants over the \textit{Main Settings}. \acs{MEDD} is the best-performing trace-only variant.}
    \label{fig:dd_comparison}
\end{figure}

Fig.~\ref{fig:detector_family_comparison} compares the selected representative detector variant per family over the \(4500\) \textit{Main Settings} payload bits. The \ac{MMSE} benchmark gives the lowest pooled \ac{BER} with \(32/4500=0.71\%\), while \ac{MEDD} follows closely with \(34/4500=0.76\%\). The remaining representatives obtain \(57/4500=1.27\%\) for \ac{DFDD}, \(78/4500=1.73\%\) for baseline-subtracted mean energy detection, \(125/4500=2.78\%\) for \ac{MF}, and \(152/4500=3.38\%\) for simplified \ac{MLSD}. Thus, \ac{MEDD} outperforms all other trace-only representatives and nearly reaches the best overall \ac{CIR}-assisted result.

For both leading detectors, all Main-Settings errors are concentrated in the medium-distance \(\qty{6}{\bit\per\second}\) condition, with \(34/300\) errors for \ac{MEDD} and \(32/300\) for \ac{MMSE}; the other listed \textit{Main Settings} are error-free over their evaluated payloads.

The ordering includes both trace-only and \ac{CIR}-assisted approaches and indicates that adding a measured \ac{CIR} does not automatically improve performance on these payload traces. Deviations between the isolated-pulse templates and the continuous live measurements may contribute to the lower performance of some assisted variants.

Viewed together with Table~\ref{tab:detector_complexity}, the small \ac{BER} difference between \ac{MEDD} and \ac{MMSE} has a useful practical interpretation. \ac{MEDD} remains in the same \(O(MN)\) symbol-window class as the basic energy and difference-detection variants, whereas the slightly lower \ac{MMSE} result requires a full-sequence benchmark solve with substantially higher computational and memory cost. This does not establish \ac{MEDD} as universally preferable, but it shows that the previous-energy scaling and offset modification is effective for this experimental dataset.
\begin{figure}[t]
    \centering
    \includegraphics[width=0.9\linewidth]{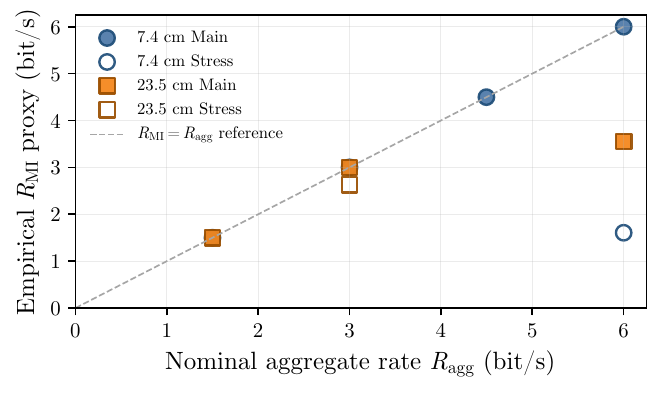}
    \caption{Empirical decision-information-rate proxy \(R_{\mathrm{MI}}\) versus nominal aggregate rate. Each point uses the decisions of a detector attaining the minimum optimized \acs{BER} at that operating point. The dashed line indicates the ideal \(R_{\mathrm{MI}}=R_{\mathrm{agg}}\) reference. Filled markers show \textit{Main Settings}; open markers show representative \textit{Stress Settings}.}
    \label{fig:mi_vs_rate}
\end{figure}
\subsection{Difference-Family Comparison}
\label{subsec:detector_difference_comparison}

Fig.~\ref{fig:dd_comparison} compares the trace-only difference variants. \ac{MEDD} is the best variant in this group with \(34/4500=0.76\%\), followed by \ac{DFDD} with \(57/4500=1.27\%\), \ac{DD} with \(61/4500=1.36\%\), and the fixed mean-normalized \ac{EDD} with \(180/4500=4.00\%\). The gain is important because the algorithmic change is small: scaling the previous energy and adding an adaptive offset improves robustness to residual tails and amplitude variation while preserving a trace-only detector structure.

The complexity comparison in Table~\ref{tab:detector_complexity} also cautions against overinterpreting the family ordering. A small \ac{BER} difference between two families can be less important than whether the detector needs only a local trace statistic, a \ac{CIR} template, an equalizer solve, or sequence-memory search.

\subsection{Measured Rate--Reliability Trade-Off}
\label{subsec:detector_main_settings}

Fig.~\ref{fig:mi_vs_rate} summarizes the detector-level decision-information proxy over the measured rate scenarios. The highest nominal aggregate rate is \(R_{\mathrm{agg}}=\qty{6}{\bit\per\second}\), corresponding to \(T_{\mathrm{sym}}=\qty{0.5}{\second}\) with three simultaneous color subchannels. For the short-distance payload traces, the \textit{Main Settings} are error-free or nearly error-free over the evaluated rate range. The \(\qty{6}{\bit\per\second}\) setting with \(T_{\mathrm{inj}}=\qty{0.2}{\second}\) is error-free over the evaluated payload and reaches the ideal empirical value \(R_{\mathrm{MI}}=\qty{6}{\bit\per\second}\).

For the medium-distance payload traces, the \(\qty{3}{\bit\per\second}\) setting with \(T_{\mathrm{inj}}=\qty{0.05}{\second}\) is also error-free over the evaluated payload and reaches \(R_{\mathrm{MI}}=\qty{2.996}{\bit\per\second}\), close to the nominal aggregate rate. The \(\qty{6}{\bit\per\second}\) setting remains visibly more challenging. The main trend is therefore not determined by \(R_{\mathrm{agg}}\) alone. Channel length, injection duration, and detector robustness jointly determine whether a nominally high-rate operating point remains useful.

\subsection{Stress Settings and Injection-Duration Trade-Off}
\label{subsec:detector_stress_settings}

Table~\ref{tab:detector_operating_points} isolates two \textit{Stress Settings} that help interpret the operating boundary. They reveal how injection duration interacts with rate, channel length, and residual tails. In the short-distance high-rate setting with \(T_{\mathrm{sym}}=\qty{0.5}{\second}\), increasing \(T_{\mathrm{inj}}\) from \(\qty{0.1}{\second}\) to \(\qty{0.2}{\second}\) improves the best operating-point \ac{BER} from \(20.72\%\) to error-free performance over the evaluated payload. In this regime, the shorter injection is too weak relative to noise, timing variation, and symbol overlap.

In the medium-distance setting with \(T_{\mathrm{sym}}=\qty{1}{\second}\), the trend is reversed: increasing \(T_{\mathrm{inj}}\) from \(\qty{0.05}{\second}\) to \(\qty{0.1}{\second}\) worsens the best operating-point result from error-free performance to a \(3.00\%\) \ac{BER}. This is consistent with a longer-injection tail/\ac{ISI} trade-off at the medium distance causing a larger temporal spread. The \textit{Stress Settings} therefore do not define a separate operating optimum, but they expose the trade-off that must be controlled when the testbed is pushed outside the \textit{Main-Settings} operating points.

\section{Conclusion}
\label{sec:conclusion}

This work extends the spectral \ac{MUMO} \ac{MC} testbed in~\cite{wietfeldEvaluationMultiMoleculeMolecular2024b} from an initial platform demonstration toward a broader experimental study of channel behavior and \ac{RX} design. The main result is a reusable flow-based setup that connects effective channel modeling with detector benchmarking on real \ac{MUMO} measurements.

\subsection{Summary of Main Findings}
\label{subsec:conclusion_lessons}

The testbed uses cyan, magenta, and yellow inks, micropump-based injection, and non-invasive spectral sensing to recover three color traces in real time. It supports simultaneous molecule-domain signaling, and after targeted parameter tuning, the detector benchmark produces zero errors over the evaluated payloads at \qty{6}{\bit\per\second} in the short-distance ($\approx\qty{8}{\centi\meter}$) channel and at \qty{3}{\bit\per\second} in the medium-distance ($\approx\qty{24}{\centi\meter}$) channel.

Repeated operation further showed that injection geometry, hydraulic pressure balance, bubble control, stable electrical connections, and accumulated microcontroller timing offsets are central to reproducible measurements.

The isolated-pulse measurements were modeled as effective end-to-end responses, where the observed \acp{CIR} combine nonideal injection, propagation, reception, amplitude scaling, and timing alignment. Both the \ac{AD}-style effective broadening model and the inner-disk Poiseuille benchmark reproduce much of the dominant arrival timing, rise, peak, and main decay across the selected distances, while late-tail mismatch remains the principal limitation. The fitted \(B_{\mathrm{eff}}\) and \(\alpha\) values are shared effective model descriptors rather than physical identifications of molecular diffusion or injection geometry.

We connected this channel characterization to communication performance through a measurement-based detector-family benchmark on continuous \ac{MUMO}-\ac{OOK} payload traces. The proposed \ac{MEDD} scheme, a previous-energy scaling and offset extension of \ac{EDD}, outperformed all other trace-only detectors with a pooled \ac{BER} of \(34/4500=0.76\%\). It differed from the best overall \ac{MMSE} benchmark by only two errors over the \(4500\) Main-Settings bits, showing that trace-only difference statistics can be highly competitive without explicit \ac{CIR} knowledge.

\subsection{Outlook Toward Multiple Access}
\label{subsec:conclusion_multiple_access}

Simultaneous C/M/Y operation can be interpreted as molecule-domain parallel signaling and, when the molecule types are assigned to different logical \acp{TX}, as a \acl{MDMA}-like configuration. A full fairness-controlled comparison of multiple access schemes should build on the presented stable payload transmission, effective color separation, matched or controlled injection-duration policies, and common \ac{RX} benchmark. Controlled spectral overlap between colors could be used to emulate molecular receptor interference. Together with existing \ac{NOMA} studies in \ac{MC}~\cite{wietfeldDBMCaNOMAlyAsynchronousNOMA2026,wietfeldDBMCNOMAEvaluatingNOMA2024}, the testbed provides an experimental basis for future multiple-access investigations.

\bibliographystyle{IEEEtran}
\bibliography{references}

\vfill

\end{document}